\documentclass[10pt, twocolumn]{IEEEtran} 

\usepackage{color}
\usepackage{array}
\usepackage{booktabs}
\usepackage{multirow}
\usepackage{amsmath}
\usepackage{amssymb}
\usepackage{graphicx}
\usepackage[numbers,numbers,sort&compress]{natbib}

\usepackage[ruled,vlined,linesnumbered]{algorithm2e}

\usepackage{graphicx}
\makeatletter

\providecommand{\tabularnewline}{\\}

\SetKwInput{KwInput}{Input}
\SetKwInput{KwOutput}{Output}

\usepackage{algpseudocode}
\usepackage{amsmath,amsfonts,amssymb}
\usepackage{array,booktabs,arydshln}
\usepackage{xcolor}
\usepackage{arydshln}
\usepackage{float} 

\usepackage[caption=false,font=footnotesize]{subfig}

\newcommand{\dcoflow}{$\mathtt{DCoflow}$}

\usepackage{tikz}
\usepackage{tcolorbox}

\@ifundefined{showcaptionsetup}{}{%
 \PassOptionsToPackage{caption=false}{subfig}}
\usepackage{subfig}
\makeatother

\newcommand{\cC}{{\cal C}}
\newcommand{\cF}{{\cal F}}

\newcommand{\cL}{{\cal L}}
\newcommand{\cO}{{\cal O}}
\newcommand{\cS}{{\cal S}}
\newcommand{\cT}{{\cal T}}

\newcommand{\hv}{\hat{v}}
\newcommand*\diff{\mathop{}\!\mathrm{d}} 

\usepackage{amsthm}
\newtheorem{theorem}{Theorem}[]
\newtheorem{lemma}[theorem]{Lemma}

\begin{document}
\title{DCoflow: Deadline-Aware Scheduling Algorithm for Coflows in Datacenter Networks}

\author{
\IEEEauthorblockN{
    Quang-Trung Luu\IEEEauthorrefmark{1},
    Olivier~Brun\IEEEauthorrefmark{1},
    Rachid~El-Azouzi\IEEEauthorrefmark{2},
    Francesco~De~Pellegrini\IEEEauthorrefmark{2},\\
    Balakrishna~J.~Prabhu\IEEEauthorrefmark{1},
    Cédric~Richier\IEEEauthorrefmark{2}\\
}\IEEEauthorblockA{\IEEEauthorrefmark{1}LAAS-CNRS, University of Toulouse, CNRS, 31400  Toulouse, France }\\
\IEEEauthorblockA{\IEEEauthorrefmark{2}CERI/LIA, University of Avignon, 84029 Avignon, France} \\
\{qtluu, brun, bala\}@laas.fr, \{rachid.elazouzi, francesco.de-pellegrini, cedric.richier\}@univ-avignon.fr
\vspace{-0.5cm}
}

\pagestyle{plain}


\maketitle

\IEEEpubidadjcol

\begin{abstract}
Datacenter networks routinely support the data transfers of distributed computing frameworks in the form of coflows, i.e., sets of concurrent flows related to a common task. The vast majority of the literature has focused on the problem of scheduling coflows for completion time minimization, i.e., to maximize the average rate at which coflows are dispatched in the network fabric. Modern applications, though, may generate coflows dedicated to online services and mission-critical computing tasks which have to comply with specific completion deadlines. In this paper, we introduce \dcoflow, a lightweight deadline-aware scheduler for time-critical coflows in datacenter networks. The algorithm combines an online joint admission control and scheduling logic and returns a $\sigma$-order schedule which maximizes the number of coflows that attain their deadlines. Extensive numerical results demonstrate that the proposed solution outperforms existing ones. 
\end{abstract}
\begin{IEEEkeywords}
Time-sensitive coflow scheduling, coflow admission control, $\sigma$-order, deadline.  
\end{IEEEkeywords}

\IEEEpeerreviewmaketitle{}

\section{Introduction
\label{sec:Introduction}}

Modern traffic engineering in datacenter networks is based on the notion of coflow originally defined  in \cite{ChowdhuryHotNet2012}. The interest for this traffic abstraction has originally been motivated by the need to capture the structure of the data exchanges occurring in distributed computing frameworks such as MapReduce or Spark \cite{dean2004mapreduce,zaharia2010spark}.
Such software frameworks rely on the so called \emph{dataflow} computing model for large-scale data processing, i.e., a distributed computing paradigm, where each intermediate computation stage is distributed over a set of nodes and its output is transferred to nodes hosting the next stage. In between two computation stages, such dataflows are producing a set of network flows traversing the datacenter fabric and are abstracted as a \emph{coflow}. A popular example of such data transfer is the \emph{shuffle phase} of Hadoop MapReduce \cite{dean2004mapreduce}. 

The customary performance metric for the data transfer phase is the makespan or the weighted coflow completion time (CCT). Minimizing the average CCT is an appropriate goal in order to increase the number of computing jobs dispatched per hour in a datacenter. The weighted CCT minimization has thus been addressed in several works, e.g., \cite{ChowdhuryHotNet2012,chowdhury2015coflow,Shaf2018,agarwal2018sincronia}. A decade's research on the problem has shed light on its complexity and several algorithmic solutions have been devised. The problem was proven {\em NP}-hard and inapproximable below a factor of $2$ by reduction to the job scheduling problem on multiple correlated machines. Near-optimal algorithms with $4$-approximation performance bounds have been proposed in the literature  \cite{agarwal2018sincronia,chowdhury2019near,Shaf2018}.
However, the context changes radically in the case of time-sensitive jobs, where the data transfer phase may be subject to strict coflow deadlines. 

Here, coflow scheduling is typically combined with admission control in order to  reduce the number of violations, i.e., the number of coflows to complete after their deadlines. The resulting problem is the coflow deadline satisfaction (CDS) problem introduced in \cite{Tseng2019}. Each coflow is subject to a completion deadline and the target is to operate joint {\em coflow admission control and scheduling} in such a way to maximize the number of admitted coflows which respect their deadlines. This problem is {\em NP}-hard as well and it is proved inapproximable within any constant factor from the optimum~\cite{Tseng2019}. 

Even though the problem has been identified quite early in the literature \cite{Chowdhury2014}, with a few exceptions, the vast majority of works on coflow scheduling have not dealt with the problem of time-sensitive coflows. On the other hand, as confirmed later in our performance analysis, even near-optimal algorithms for CCT minimization may fail to respect the coflow deadline. In reality, the notion of time-sensitive coflows has become pervasive in the way how modern datacenters operate as distributed networks. In fact, not only computing frameworks are often tasked with time-sensitive jobs: modern web and mobile applications are implemented using microservice architectures, so that the users requests issued to an application may activate hundreds or thousands of services from as many servers to retrieve the users data. The last incoming bit of data, i.e., the CCT of this batch of flows, determines the lag to the service response, and large delays degrade the quality of experience.


In this paper, we address the problem of maximizing the Coflow Acceptance Rate (CAR), in which each coflow is subject to a completion time deadline. In principle, one can solve this problem by formulating a suitable Mixed Integer Linear Program (MILP). However, in datacenters with tens of thousands of coflows~\cite{chowdhury2015coflow}, techniques based on MILPs or their relaxations may be not viable. For designing scalable algorithms, the main idea appearing in many research works is to schedule coflows using a priority order of coflows.  Once an ordering, denoted as $\sigma$, is determined,  it
is enough to adopt a work-conserving transmission policy.  We focus on $\sigma$-order schedulers since they offer a key implementation advantage: at the level of rate control, any work conserving preemptive dynamic rate allocation is allowed as long as it is compatible with the input coflow priority (the maximal performance loss within said rate allocation policies is bounded by a factor of $2$ \cite{agarwal2018sincronia}). For instance, using fixed coflow priorities under DiffServ satisfies the definition of a $\sigma$-order scheduler. On the other hand, in order to avoid per-flow rate control, commercial switches have built-in priority queues and per-flow tagging can be used to prioritize active coflows. In principle, this permits to perform a greedy rate allocation which is compatible with a target $\sigma$-order. The exact mapping from a coflow $\sigma$-order to the switch priority queuing mechanism (and the inevitable limitations of legacy hardware therein) while an interesting subject, is out of the scope of the present paper.

\vspace{0.2cm}
\noindent\textit{Contributions.} This paper proposes a lightweight method to perform coflow scheduling under deadlines. The proposed solution provably outperforms existing ones in the literature and does not rely on the solution of a linear program. In particular, we propose first a baseline offline admission control policy which is combined with a scheduler drawn in the class of 
$\sigma$-order coflow schedulers \cite{agarwal2018sincronia}. The output of the algorithm is an order of priority restricted to the set of admitted coflows. The algorithm is hence extended to perform joint admission control and scheduling in the online scenario, when coflows are generated at runtime at unknown release times. Through extensive numerical experiments on a wide variety of scenarios, we show that our algorithm systematically outperforms the ones currently available in the literature. These experiments are performed on both offline and online setting using both synthetic traces and real traces obtained from the Facebook data \cite{Chowdhury2014}. The main observation is that, under higher workloads (i.e., when the acceptance ratio is lower), our algorithm outperforms significantly 
the existing ones. 
Thus, it proves robust to workload variations as well to the type of data set used to generate the coflows.




The rest of the paper is organized as follows. Sec.~\ref{sec:Problem-Statement} describes the general problem tackled in the paper and the coflow ordering models, whereas  Sec.~\ref {sec:heuristic} describes the proposed algorithms. Numerical results are then provided in Sec.~\ref{sec:Evaluation}. In Sec.~\ref{sec:Related-Work}, we describe the literature on deadline-aware coflow scheduling. Concluding remarks and future research directions are given in Sec.~\ref{sec:Conclusion}.

\section{Problem Statement and Theoretical Analysis 
\label{sec:Problem-Statement}}

In this section, we present the system model and formulate the acceptance rate maximization problem for a given input set of coflows. Table~\ref{tab:Prob:Notations} summarizes the main notations used throughout the paper.
The datacenter network (or datacenter fabric) is represented as a \textit{Big-Switch} model \cite{Chowdhury2014}, a non-blocking switch whose ingress (egress) port $\ell$ has capacity $B_{\ell}$ equal to the corresponding bandwidth inbound (outbound) capacity to connect servers to the top-of-rack (ToR) switch. Due to large bisection capacity and customary usage of load balancing, in fact, traffic congestion is typically observed only at the rack access ports leading to the ToR switches.

We consider a batch of $N$ coflows $\cC = \{1, 2, ..., N\}$. A coflow is a collection of flows, in which each flow represents a shuffle connection over a pair of fabric ingress-egress ports. Denote $\cF_{k}$ as the set of flows of coflow $k \in\cC$ and assume that the volume $v_{k,j}$ of each flow $j \in \cF_k$ is known. For the sake of clarity, we suppose that all coflows arrive at the same time, i.e., their release time is zero. Also, each coflow $k$ is subject to a completion deadline $T_k$. Similarly, we let $\cF_{\ell,k}$ be the set of flows in $\cF_{k}$ which uses port $\ell \in \cL$ either as ingress port or as egress port. The total volume of data sent by coflow $k$ on port $\ell$  is then given by  $\hv_{\ell,k}=\sum_{j \in \cF_{\ell,k}} v_{k,j}$. Let $p_{\ell,k}$ be the transfer completion time in isolation, i.e., at full rate, of coflow $k$ at port $\ell$, which is given by $p_{\ell,k} = \hv_{\ell,k} / B_\ell$. 
Let $r_{k,j}(t) \in \mathbb {R}_{+}$ be the rate allocated to flow $j\in {\cal F}_k$ at time $t$. The CCT of coflow $k$, denoted as $c_k$, thus writes
\begin{equation}
\hspace{-0.2cm} c_k = \max_{j\in {\cal F}_k}  \frac{v_{k,j}}{\bar r_{k,j}}, \mbox{ where}~  \bar{r}_{k,j} =\frac{1}{CT_{k,j}} \int_0^{CT_{k,j}} r_{k,j} (t) \diff{t},
\end{equation}
and $\bar r_{k,j}$ is the average rate of flow $j$ through its lifetime and  $CT_{k,j}$ is its completion time. This quantity rules the dispatching time for the volume traversing the so-called coflow bottleneck which determines the CCT of the said coflow. Hence a coflow $k$ satisfies the deadline when the last flow of  the coflow finishes before $T_k$, i.e., $c_k\leq T_k$. 

Let $z_{k} \in \{0,1\}$ be an indicator of whether coflow $k$ finishes before $T_k$. Maximizing the number of coflows meeting their deadline corresponds to maximizing the sum of $z_k$ subject to the constraint of bandwidth capacity at ingress and egress ports. The CAR maximization problem can be formulated as 


\vspace{-0.5cm}

\begin{align}
\underset{r}{\mathrm{max}}\enskip & \sum_{k\in\mathcal{C}}  z_{k} \tag{P1}\label{prob:CAR}\\
\mathrm{s.t.}\enskip 
& \sum_{k \in \mathcal{C}} \sum_{j\in\mathcal{F}_{k,l}} r_{k,j}(t)\leq  B_{\ell}, \quad \forall \ell \in \mathcal{L}, \forall t \in \mathcal{T}, \label{prob:CAR-1} \\
& \int_0^{T_k} r_{k,j} (t) \diff{t} \geq  v_{k,j} z_k, \quad \forall j \in \mathcal{F}_k, \forall k \in \mathcal{C}, \label{prob:CAR-2}
\end{align}
Constraint \eqref{prob:CAR-1} expresses that, at any instant $t$ within the time horizon $\cT$, the total rate that port $\ell$ assigns to flows cannot exceed its capacity $B_{\ell}$. Constraint \eqref{prob:CAR-2} ensures that the data of flows of each \textit{accepted} coflow $k$ should be completely transmitted before the deadline $T_k$. 

\begin{lemma}[Proposition 1 in \cite{Tseng2019}]
There exists a polynomial time reduction of the CAR problem \eqref{prob:CAR} to the problem of minimizing the number of late jobs in a concurrent open shop \cite{Lin2007}. Hence, the CAR problem is \textit{NP}-hard. 
\end{lemma}

\vspace{-1cm}

\begin{center}
\begin{table}[t]
\caption{Main notations. \label{tab:Prob:Notations}}
\footnotesize
\centering{}%
\begin{tabular}{cl}
\toprule 
{Symbol} & {Description} \tabularnewline
\cmidrule[0.4pt](lr{0.12em}){1-1}%
\cmidrule[0.4pt](lr{0.12em}){2-2}%
$\cL$ & set of fabric ports\tabularnewline
$M$ & number of machines, $M = |\cL|/2$\tabularnewline
$B_{\ell}$ & available bandwidth of port $\ell\in\mathcal{L}$\tabularnewline
$\cC$ & set of coflows. $\cC$ has cardinality of $N$\tabularnewline
$\sigma$ & scheduling order of coflows, $\sigma=\left\{ \sigma_{1},\cdots,\sigma_{N-1},\sigma_{N}\right\} $\tabularnewline
$T_{k}$ & deadline of coflow $k$\tabularnewline
$\cF_k$ ($\cF_{\ell,k}$) & set of flows of coflow $k$ (that use port $\ell$) \tabularnewline
$v_{k,j}$ & volume of flow $j \in \cF_k$ of coflow $k$\tabularnewline
$\hv_{\ell,k}$ & total volume transmitted by coflow $k$ on port $\ell$\tabularnewline
$p_{\ell,k}$ & processing time of coflow $k$ on port $\ell$\tabularnewline
$c_{k}$ ($c_{\ell,k}$) &  completion time of coflow $k$ on port $\ell$ \tabularnewline
${CT}_{k,j}$ & completion time of flow $j$ of coflow $k$ \tabularnewline
\bottomrule
\end{tabular}
\end{table}
\end{center}

\vspace{-0.5cm}

\subsection{Motivating Example}
\label{subsec:Motivating-Example}
 

We will take $\mathtt{CS\text{-}MHA}$~\cite{Luo2016} as our starting point to address the CAR problem. $\mathtt{CS\text{-}MHA}$  has introduced a new direction to solve the scheduling problem by means of a static coflow prioritisation. The prioritisation is used in order to approximate the solution of the coflow scheduling problem that maximizes the CAR. $\mathtt{CS\text{-}MHA}$ computes the scheduling orders at each port using the Moore-Hodgon's algorithm which also determines the set of admitted coflows at each port $\ell\in{\mathcal{L}}$.  Since different ports could have different  sets of admitted coflows, a coflow is admitted if it is admitted at all ports. For all rejected coflows, a second round is applied to check if some rejected coflows can actually satisfy their deadline. The algorithm selects a coflow with the minimum bandwidth required at the bottleneck port since it is more likely to catch up with its deadline.

\begin{figure}[h]
\centering
\includegraphics[width=0.5\columnwidth]{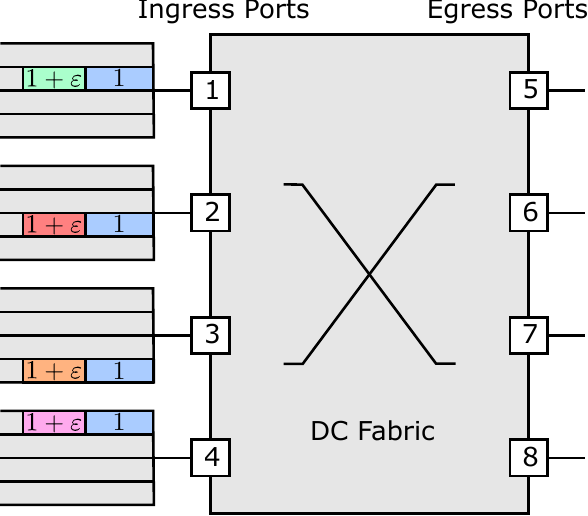}
\caption{Motivating example with a Big-Switch fabric composed of $4$ ingress/egress ports connecting to $4$ machines. Flows in ingress ports are organized by destinations and are color-coded by coflows. The example has $5$ coflows. Coflow $C_1$ (blue) has $4$ flows, with each ingress port sending $1$ units of data to one egress port: its deadline is $1$;
coflows $C_2$ (green), $C_3$ (red), $C_4$ (orange) and $C_5$ (purple) have a single flow, each sending $(1+\varepsilon)$ unit of data. The deadline of these coflows is $2$.  \label{fig:Example}}
\end{figure}

The simple example depicted in Fig.~\ref{fig:Example} illustrates the limitations of  $\mathtt{CS\text{-}MHA}$. It uses the standard \emph{Big-Switch} model to abstract a datacenter fabric: the example will be used as a running example throughout the paper. The instance contains $5$ coflows: $C_{1}$ has $4$ flows and $C_{2}$, $C_{3}$, $C_{4}$ and $C_{5}$ each have one flow. To ease the presentation, the flows are organised in virtual output queues at the ingress ports. The virtual queue index represents the flow output port, modulo the number of machines. The numbers on the flows' representations correspond to their \textit{normalized} volumes. All fabric ports have the same \textit{normalized} bandwidth of $1$.

At the first iteration, $\mathtt{CS\text{-}MHA}$ computes the scheduling order  on each port using Moore-Hodgson algorithm \cite{Moore1968}, which is based on the \textit{Earliest Due Date} rule and is known to minimize the number of missed deadlines on a single machine (or \textit{port} in the coflow context). Since coflow $C_{1}$ uses all ports and has the smallest deadline ($T_{1} = 1$), then at each port, $C_{1}$ will be scheduled first. As consequence, all other coflows are rejected since they cannot satisfy their deadline when scheduled after $C_{1}$. Given that coflow scheduling, the CAR is $\frac{1}{5}$. However, an optimal scheduling solution would be $\{C_{2}, C_{3}, C_{4}, C_{5}, C_{1}\}$ or any combination that has coflow $C_{1}$ as the last one to be scheduled. The latter scheduler attains a CAR of $\frac{4}{5}$. From this example, we can see that $\mathtt{CS\text{-}MHA}$ can be made arbitrarily worse compared to the optimal solution. 

The above example can easily be extended to show that in the worst-case scenario, the CAR of  $\mathtt{CS\text{-}MHA}$ can be made arbitrarily small (as close to $0$ as needed). Consider $M$ machines, coflow $C_1$ that uses all ports, and coflows $C_2,\hdots, C_M$ with one flow each. The other parameters remain the same as above. It can be seen that the CAR obtained using $\mathtt{CS\text{-}MHA}$  and $\mathtt{DCoflow}$ are $\frac{1}{M}$, $\frac{M-1}{M}$, respectively.  In this setting, the CAR obtained using $\mathtt{CS\text{-}MHA}$  is close to zero when $M$ is high, while  using  $\mathtt{DCoflow}$, the CAR is close to one.

The key observation is that $\mathtt{CS\text{-}MHA}$ neglects the impact that a coflow may have on other coflows on multiple ports. Indeed, a coflow causing multiple deadlines to be missed should have lesser priority, even when its deadline is the earliest. When this is neglected, the coflow ordering is misjudged thus degrading the CAR. In this work, we start from this observation and propose a new $\sigma$-order scheduler, called $\mathtt{DCoflow}$.

\vspace{-0.3cm}
\subsection{Coflow Ordering 
\label{subsec:Coflow-Ordering}}

Problem~\eqref{prob:CAR} implicitly depends on the CCT. An alternative approach to maximize the CAR is to {\em order} the coflows in some appropriate way, and then to leverage the priority forwarding mechanisms of the underlying transport network \cite{Agarwal2018}.


In this approach, 
once such an ordering $\sigma$ is determined, it is enough to adopt work-conserving transmission policies and use forwarding priorities in such a way that a flow $j \in \cF_{\sigma_{k}}$ is blocked if and only if either its ingress or egress port is busy serving a flow $j' \in  \cF_{\sigma_{k'}}$ for some $k' < k$, i.e., a flow of a coflow with higher priority according to $\sigma$. Such a flow scheduling is called \emph{$\sigma$-order-preserving}. 

The formulation of Problem~\eqref{prob:CAR} can be transformed into an Integer Linear Program (ILP) as follows. 
We define the binary variable $\delta_{i,k}$ as $1$ if coflow $i$ has higher priority than coflow $k \neq i$, and $0$ otherwise. The ordering of coflows can then be modeled using the following standard disjunctive constraints
\begin{eqnarray}
\delta_{k,k'} + \delta_{k',k} & = & 1, \quad \forall k,k' \in \cC, \label{eq:order-disjunct-1} \\
\delta_{k,k'} + \delta_{k',k"} + \delta_{k",k} & \leq & 2, \quad \forall k,k',k" \in \cC. \label{eq:order-disjunct-2}
\end{eqnarray}

It should be clear that the ordering $\sigma$ can easily be derived from the variables $\{ \delta_{k,k'} \}_{k,k' \in \cC}$. This ordering should be such that as many coflows as possible are accepted, that is, $\sum_{i \in \cC} z_k$ is maximized. A central difficulty here is to compute the completion time of a coflow, which stems from the fact that data transmissions on the various ports are not independent: the transmission of a flow may be blocked until the ingress port becomes available even if the egress port is idle, and vice versa. However, assuming that the ports are independent, we can obtain a lower bound on the completion time of coflow $k$ on port $\ell$ as follows
\begin{equation} \label{eq:lower-bound-CCT}
c_{\ell,k} \geq \sum_{k' \neq k} p_{\ell,k'} \delta_{k',k} z_{k'} + p_{\ell,k}, \quad \forall \ell \in \cL, k \in \cC.
\end{equation}
This lower bound assumes that the transmission of coflow $k$ on port $\ell$ can start as soon as all flows of all coflows $k'$ scheduled before $k$ have been transmitted on port $\ell$. Note that the above constraint is not linear due to the product $\delta_{k',k} z_{k'}$. However, it can easily be linearized by introducing binary variables $y_{k',k}$ satisfying the constraints 
\begin{equation}
y_{k',k} \leq z_{k'}; 
\quad y_{k',k} \leq \delta_{k',k}; 
\quad y_{k',k} \geq z_{k'}+\delta_{k',k} - 1, \label{eq:y1} 
\end{equation}
\eqref{eq:lower-bound-CCT} thus becomes
\begin{equation} \label{eq:lower-bound-CCT-new}
c_{\ell,k} \geq \sum_{k' \neq k} p_{\ell,k'} y_{k',k} + p_{\ell,k}, \quad \forall \ell \in \cL, k \in \cC.  
\end{equation}
A lower bound on the completion time of coflow $k$ is then obtained as $c_k = \max_{\ell \in \cL} c_{\ell,k}$, and this coflow can meet its deadline only if $c_k \leq T_k$. The deadline constraint can then be described as
\begin{equation} \label{eq:deadline-constraint}
c_{\ell,k} \leq T_k z_k, \quad \forall \ell \in \cL, k \in \cC.  
\end{equation}
We can thus obtain an upper bound on the number of accepted coflows by solving the following ILP, 
\begin{equation} \label{prob:ILP}
\mathrm{max}\enskip \sum_{k\in\mathcal{C}} z_{k}\tag{P2}, \quad
\mathrm{s.t.}\enskip (\ref{eq:order-disjunct-1}, \ref{eq:order-disjunct-2}, \ref{eq:y1}, \ref{eq:lower-bound-CCT-new}, \ref{eq:deadline-constraint}).\nonumber
\end{equation}
It is easy to verify that Problem~\eqref{prob:ILP} is \textit{NP}-hard and so that it may become unfeasible to find an optimal solution in large-scale datacenter networks. In the next section, we hence propose an efficient heuristic algorithm to determine the order in which coflows must finish in order to maximize the coflow acceptance rate.

\section{$\sigma$-Order Scheduling with $\mathtt{DCoflow}$ 
\label{sec:heuristic}}

In this section, we present $\mathtt{DCoflow}$, an algorithm to solve the problem of joint coflow admission control and scheduling. Given a list of $N$ coflows, it provides a permutation $\sigma=(\sigma_1,\sigma_2,..,\sigma_N)$ of these coflows, with the aim of maximizing the coflow acceptance rate. We observe that, when a coflow is served according to the corresponding $\sigma$-order-preserving schedule, lower-priority coflows will be impacted. It is hence possible that some of these lower-priority coflows will not respect their deadline. Hence our objective is to evaluate the impact of a scheduled coflow with respect to the original CAR problem. The problem is combinatorial in nature, since there are $N!$ possible coflow orderings. Hence, we propose a new approach that uses the formulation of Problem~\eqref{prob:ILP}  by deriving a necessary condition  satisfied by all feasible coflow orders solving Problem~\eqref{prob:ILP}. 

Consider a feasible solution to Problem~\eqref{prob:ILP} and  a subset $\cS \subseteq \cC$ of coflows. Given a coflow $k \in \cC$, let $\cS_k^-=\left \{ k' \in \cS \ : \ \delta_{k',k} = 1\right \}$ be the set of coflows in $\cS$ which are scheduled before $k$. Condition \eqref{eq:lower-bound-CCT} may be rewritten as
\begin{equation}
c_{\ell,k} \geq p_{\ell,k} + \sum_{k' \in \cS_k^-} p_{\ell,k'} z_{k'},
\label{eq:parallel-inequalities-1}
\end{equation}
which implies
\begin{equation}
c_{\ell,k} p_{\ell,k} z_k \geq p_{\ell,k}^2 z_k^2 + p_{\ell,k} z_k \sum_{k' \in \cS_k^-} p_{\ell,k'} z_{k'}.
\label{eq:parallel-inequalities-2}
\end{equation}

Using the inequality $T_k \geq c_{\ell,k} z_k $ and summing over all coflows $k \in \cS$, we obtain
\begin{small}
\begin{align} 
\sum_{k\in\mathcal{S}}p_{\ell,k}T_{k} \geq & \sum_{k\in\mathcal{S}}c_{\ell,k}p_{\ell,k}z_{k} \geq   \sum_{k\in\mathcal{S}}\left(p_{\ell,k}z_{k}\right)^{2} + \hskip-4mm \sum_{k\in\mathcal{S}, k'\in\mathcal{S}_{k}^{-}} \hskip-4mm
p_{\ell,k}p_{\ell,k'}z_{k}z_{k'}\nonumber \\
 & \hskip-19mm=\frac{1}{2}\sum_{k\in\mathcal{S}}\left(p_{\ell,k}z_{k}\right)^{2}+\nonumber \frac{1}{2}\left[\sum_{k\in\mathcal{S}}\left(p_{\ell,k}z_{k}\right)^{2}+2\sum_{k\in\mathcal{S}}p_{\ell,k}z_{k} \hskip-2mm \sum_{k'\in\mathcal{S}_{k}^{-}}p_{\ell,k'}z_{k'}\!\right]\nonumber \\
&\hskip-19mm = \frac{1}{2}\sum_{k\in\mathcal{S}}\left(p_{\ell,k}z_{k}\right)^{2}+\frac{1}{2}\left(\sum_{k\in\mathcal{S}}p_{\ell,k}z_{k}\right)^{2}.\label{eq:parallel-inequalities}
\end{align}
\end{small}
We thus conclude that any feasible solution to Problem~\eqref{prob:ILP} satisfies the condition $\sum_{k \in \cS} p_{\ell,k} T_k \geq f_\ell(\cS)$ for any subset $\cS \subseteq \cC$ of \emph{accepted} coflows, where $f_\ell(\cS)=\frac{1}{2}  \sum_{k \in \cS} p_{\ell,k}^2 +  \frac{1}{2}  \left ( \sum_{k \in \cS}  p_{\ell,k} \right )^2$. These conditions are the so-called \textit{parallel inequalities} and provide valid inequalities for the concurrent open shop problem \cite{Mastrolilli2010}. Note that they do not depend on the ordering of the coflows. 

We shall use the parallel inequalities to determine the coflows that should not be admitted. Consider a solution to Problem~\eqref{prob:ILP} and assume that, in this solution, there exists a subset $\cS$ of accepted coflows (i.e., $z_k=1$ 
for all $k \in \cS$) such that $\sum_{k \in \cS} p_{\ell,k} T_k < f_\ell(\cS)$ for at least one port $\ell \in \cL$. This implies that this solution is not feasible and can only become feasible by rejecting a coflow $k'$ among the accepted coflows using port $\ell$. We choose this coflow $k'$ so as to minimize the quantity $f_\ell(\cS \setminus \{ k' \}) - \sum_{k \in \cS \setminus \{ k' \}} p_{\ell,k} T_k$, in the hope that it becomes negative. Observe that 

\vspace{-0.5cm}
\begin{small}
\begin{align}
f_{\ell}\left(\cS \right) & \! =\!   \left [ \frac{1}{2}p_{\ell,k'}^{2}\! +\! \frac{1}{2}\! \sum_{k \in \cS \setminus \{ k' \} } p_{\ell,k}^{2} \right ] \! +\!  \frac{1}{2}\!  \left ( p_{\ell k'}+\! \sum_{k \in \cS \setminus \{ k' \} }p_{\ell k} \right )^{2} \nonumber \\
 & =  f_{\ell} \left ( \cS \setminus \{ k' \} \right ) + p_{\ell, k'} \sum_{k \in \cS} p_{\ell,k}, \label{eq:fls_split}
\end{align}
\end{small}
from which it follows that
\begin{small}
\begin{align}
f_{\ell}\left(\cS \setminus \{ k' \} \right)- \sum_{k \in \cS \setminus \{ k' \} } p_{\ell,k}T_k  & =
f_{\ell} \left(\cS \right) - \sum_{k \in \cS} p_{\ell,k} T_k  \nonumber \\
 &  + p_{\ell, k'} \left ( T_{k'}-\sum_{k \in\cS} p_{\ell,k} \right )\label{eq:fls_constraint_split}
\end{align}
\end{small}
The above relation will be used to define our coflow admission control algorithm. In general, it suggests possible heuristics to remove a coflow $k'$ from $\cS$ in order to satisfy the parallel inequalities. For instance, one such heuristics is to minimize the quantity 
\begin{equation}
\Psi_{\ell,k'} := p_{\ell,k'} \left ( T_{k'}-\sum_{k \in \cS} p_{\ell,k} \right ).  
\end{equation}
In fact, the term $\sum_{k \in \cS} p_{\ell,k} - T_{k'}$ represents how much the completion time of coflow $k'$ on port $\ell$ exceeds its deadline, assuming that it is scheduled latest. Hence, we could reject coflow $k'$ with both large processing time $p_{\ell,k'}$ on port $\ell$ and large deadline violation. More admission rationales based on \eqref{eq:fls_constraint_split} will be detailed in  describing $\mathtt{DCoflow}$, whose pseudocode is reported in Algorithm~\ref{algo:DCoflow}. Generally, $\mathtt{DCoflow}$ takes a list of unsorted coflows and provides as output the scheduling order of accepted coflows. It works in rounds and at each round, it either accepts a coflow or it rejects one.

\begin{algorithm}[t]
\footnotesize
\caption{$\mathtt{DCoflow}$ \label{algo:DCoflow}}

\BlankLine

$\cS=\left \{ 1,2,\ldots, N \right \}$; \Comment{initial set of unscheduled
coflows}

$\sigma = \varnothing;$ \Comment{initial scheduling order}

$\sigma^{\star} = \varnothing;$ \Comment{initial set of pre-rejected coflows}

$n = N$;  \Comment{round counter}

\While{ $\cS \neq \varnothing$ }{

    
    $\ell_b = \underset{\ell \in {\cL}}{\mathrm{arg\,max}}\,  \sum_{k \in \cS} p_{\ell, k}$; \Comment{bottleneck port}
    

	$\cS_{\ell_b} = \left\{ k \in {\cS} : p_{\ell_b, k} > 0 \right\}$; \Comment{set of coflows using $\ell_b$}

	{\it \# Coflows that can finish in time when scheduled last}

	$\cS^d_{\ell_b}\leftarrow \{j\in \cS_{\ell_b} |  \sum_{k \in \cS_{\ell_b}} p_{\ell_b,k} \leq T_j\}$; 

	\eIf{$\displaystyle \cS^d_{\ell_b} \not = \varnothing$}
	{
		
        $k_n = \underset{k \in {\cC}}{\mathrm{arg\,max}} \, T_k $; \Comment{admit coflow with largest deadline}
        
		 
		$\sigma_{n} = k_n$; \Comment{append $k_n$ to $\sigma$} 

		$\cS = \cS \setminus \left\{ k_n \right\} $; \Comment{remove $k_n$ from $\cS$} 

	}{
		
		$k^{\star} = \mathtt{RejectedCoflow}(\cS)$; \Comment{select a coflow to reject}
		
		$\sigma_{n} = k^{\star}$; \Comment{append $k^{\star}$ to $\sigma$}; 
		
		$\sigma^{\star} = \sigma^{\star} \cup \left\{ k^{\star} \right\}$; \Comment{append $k^{\star}$ to $\sigma^{\star}$}

		$\cS = \cS \backslash \left\{ k^{\star} \right\}$; \Comment{remove $k^{\star}$ from $\cS$} 

	} 
	
	$n = n-1$; \Comment{update the round index} 

} 
$\sigma = \mathtt{RemoveLateCoflows} \left( \sigma, \sigma^{\star} \right)$; 

\KwRet $\sigma$;  \Comment{final scheduling order}

\vspace{0.1cm}

\SetKwFunction{FMain}{RemoveLateCoflows} 

\SetKwProg{Fn}{Function}{:}{} 


\Fn{\FMain{\textit{$\sigma$, $\sigma^{\star}$}}}{ 





\While{ $\sigma^{\star}\neq\varnothing$ }{


$k^{\star}\leftarrow\underset{k}{\mathrm{arg\,min}}\left\{ \sigma_{k}\in\sigma^{\star}\right\} $; \Comment{first coflow in $\sigma^{\star}$}


$c_{\sigma_{k^{\star}}}\leftarrow\mathtt{evalCCT}\left(\left\{ \sigma_{k}\right\} _{k\in\left[1,k^{\star}\right]}\right)$; \Comment{CCT of coflow $\sigma_{k^{\star}}$}

\If{ $c_{\sigma_{k^{\star}}} > T_{\sigma_{k^{\star}}}$}{
$\sigma\leftarrow\sigma\backslash\left\{ \sigma_{k^{\star}}\right\} $;
} 
$\sigma^{\star}\leftarrow\sigma^{\star}\backslash\left\{ \sigma_{k^{\star}}\right\} $;
} 
\KwRet $\sigma$;  
} 
\end{algorithm}

$\mathtt{DCoflow}$ starts by computing the total completion time at each port and finds the bottleneck $\ell_b$, i.e., the port with the largest completion time. Having this, it determines $\mathcal{S}^d_{\ell_b}$, the set of coflows active on the bottleneck port $\ell_b$ (line $8$) which complete before their deadline if scheduled last on $\ell_b$. 
If $\mathcal{S}^d_{\ell_b}$ is not empty (Lines 10--13), it selects a coflow $k_n \in \mathcal{S}^d_{\ell_b}$ that has the largest deadline in $\mathcal{S}^d_{\ell_b}$. 
If $\mathcal{S}^d_{\ell_b}$ is empty (Lines 14--18), i.e., no coflow respects its deadline when scheduled last on the bottleneck (Line $13$), the algorithm selects a coflow to be removed using function $\mathtt{RejectedCoflow}$, whose aim is to comply with parallel inequalities according to \eqref{eq:fls_constraint_split}. We consider two variants, each of which corresponds to a slightly different criterion to select the candidate coflow $k^{\star}$ by taking into account its weight. 


The first variant, namely $\mathtt{DCoflow\_v1}$, 
finds the candidate $k^{\star}\in\mathcal{S}_{b}$ and every port $\ell$
used by $k$ where it fails to meet the deadline (that is, $\Psi_{\ell, k}<0$),
\begin{equation}
k^{\star}=\underset{k\in\mathcal{S}_{b}}{\mathrm{arg\,min}}\,{\displaystyle \Big( \sum_{\ell:\Psi_{\ell, k}<0}}\Psi_{\ell, k}\Big).
\end{equation}
The second variant, namely $\mathtt{DCoflow\_v2}$, finds
the candidate $k^{\star}\in\mathcal{S}_{b}$ and every port $\ell$
used by $k$ that has at least $\gamma$ times the congested level
of the bottleneck, \textit{i.e}., $\ell:\sum_{j}p_{\ell j}\geq\gamma\sum_{j}p_{bj}$,
\begin{equation}
k^{\star}=\underset{k\in\mathcal{S}_{b}}{\mathrm{arg\,min}}\,{\displaystyle \Big(  \sum_{\ell:\sum_{j}p_{\ell j}\geq\gamma\sum_{j}p_{bj}}}\Psi_{\ell k}\Big).
\end{equation}
Once the initial scheduling order $\sigma$ is obtained, $\mathtt{DCoflow}$ uses the function $\mathtt{RemoveLateCoflows}$ to estimate the CCT and remove from $\sigma$ the coflows that do not satisfy the deadline constraint (Line $20$). 
Briefly, $\mathtt{RemoveLateCoflows}$ considers each pre-rejected coflow $k^{\star}$ in $\sigma^{\star}$ (Line $24$). It calls the function $\mathtt{evalCCT}$ to evaluate the CCT of $k^{\star}$, given all the coflows in $\sigma$ that are scheduled before $k^{\star}$ (Line $25$). If $k^{\star}$ cannot meet its deadline, i.e., $c_{\sigma_{k^{\star}}} > T_{\sigma_{k^{\star}}}$, it will be removed permanently from $\sigma^{\star}$ and $\sigma$. $\mathtt{RemoveLateCoflows}$ performs a swipe on the admitted coflows (as in \cite{Luo2016}) and iteratively removes coflows in $\sigma$ that belong to $\sigma^{\star}$ until the estimated CCT $c_{\sigma_{k}}$ of each coflow $\sigma_{k} \in \sigma$ is of at most $T_{\sigma_{k}}$.

It is important to note that the final solution yielded by $\mathtt{DCoflow}$ does not guarantee all coflows in $\sigma$ to satisfy their deadlines. In Sec.~\ref{sec:Evaluation}, the prediction error of $\mathtt{DCoflow}$  indicates the gap between the estimated CAR and the actual CAR after applying resource allocation.  


\vspace{0.2cm}
{\noindent\em Example.} 
To illustrate the difference between $\mathtt{DCoflow}$  and  $\mathtt{CS\text{-}MHA}$, we consider again the example  illustrated in Fig.~\ref{fig:Example}. Table \ref{tab:Example} shows the execution of $\mathtt{DCoflow\_v1}$ on that example. At the first step, $\mathtt{DCoflow\_v1}$ chooses bottleneck ingress port $1$, which is used by coflows $C_{1}$ and $C_{2}$. It computes $\overline{\Psi}_{ k} = \sum_{\ell:\Psi_{\ell, k} < 0 } \Psi_{\ell, k}$ for both coflows  and  selects the coflow with smallest $\bar\Psi_{ k}$ (in this case, $C_{1}$) to be scheduled last. Since all unscheduled coflows do not share any port in the fabric, any ordering of remaining coflows gives the same average CAR.  Given the final schedule, $\mathtt{DCoflow}$ obtains $\frac{4}{5}$  as the average CAR, which is the optimal solution.

\begin{table}[t]
\caption{Execution of $\mathtt{DCoflow}$ on the example of Fig.~\ref{fig:Example}. \label{tab:Example}}
\centering
\begin{tabular}{
>{\raggedright}m{0.38\columnwidth}
>{\centering}p{0.05\columnwidth}
>{\centering}p{0.37\columnwidth}
}
\toprule 
Unscheduled coflows (set $\mathcal{S}$) & $\ell_{b}$ & 
$ \left\{ 
\overline{\Psi}_{1},
\overline{\Psi}_{2},
\overline{\Psi}_{3},
\overline{\Psi}_{4},
\overline{\Psi}_{5}
\right\} $ \tabularnewline
\cmidrule[0.4pt](lr{0.12em}){1-1}%
\cmidrule[0.4pt](lr{0.12em}){2-2}%
\cmidrule[0.4pt](lr{0.12em}){3-3}%
$\mathcal{S}=\left\{ \boldsymbol{C_{1}},\boldsymbol{C_{2}},C_{3},C_{4},C_{5}\right\} $ & 1 & $\left\{ -4\left(1+\varepsilon\right),-\varepsilon,\,\cdot\,,\,\cdot\,,\,\cdot\,\right\} $\tabularnewline
$\mathcal{S}=\left\{ \boldsymbol{C_{2}},C_{3},C_{4},C_{5}\right\} $ & 1 & $\left\{ \,\cdot\,,0,\,\cdot\,,\,\cdot\,,\,\cdot\,\right\} $\tabularnewline
$\mathcal{S}=\left\{ \boldsymbol{C_{3}},C_{4},C_{5}\right\} $ & 2 & $\left\{ \,\cdot\,,\,\cdot\,,0,\,\cdot\,,\,\cdot\,\right\} $\tabularnewline
$\mathcal{S}=\left\{ \boldsymbol{C_{4}},C_{5}\right\} $ & 3 & $\left\{ \,\cdot\,,\,\cdot\,,\,\cdot\,,0,\,\cdot\,\right\} $\tabularnewline
$\mathcal{S}=\left\{ \boldsymbol{C_{5}}\right\} $ & 4 & $\left\{ \,\cdot\,,\,\cdot\,,\,\cdot\,,\,\cdot\,,0\right\} $\tabularnewline
\bottomrule
\end{tabular}
\end{table}

\vspace{0.2cm}
{\noindent\em Online Implementation.}
\dcoflow\ can be also be run online, when coflows arrive sequentially and possibly in batches. For this, define $f$ to be the frequency of updates, i.e., instants at which \dcoflow\ recomputes a schedule. The updates can be performed either at arrival instants of coflows (in which case we set $f=\infty$) or periodically with period $1/f$. We assume that the scheduler knows the volumes of the flows of each arrived coflow. However, it neither knows the volumes nor the release times of future coflows. 

At each update instant, \dcoflow\ recomputes the scheduling order for coflows currently available in the network. These include the ones that were scheduled in the previous scheduling instants and have not yet finished; the ones that were rejected in the previous scheduling instants but whose deadline has not yet expired and the arrivals during the update interval. \dcoflow\ calculates the new order for this set of coflows based on the remaining volumes of the flows and not on the original volumes. Note that it is assumed that coflows can be preempted \cite{chowdhury2015coflow}. This process is repeated at each update instant.

\vspace{0.2cm}
{\noindent\em Complexity Analysis.} The complexity of $\mathtt{DCoflow}$ is $\cO(N^2)$. Specifically, in $\mathtt{DCoflow}$, the values $\sum_{i \in \cS} p_{\ell,i}$ and  $\Psi_{\ell, x}$ of each remaining coflow $x$ can be pre-stored by calculating them at a cost $\cO(NL)$, where $L = |\cL|$ is the number of ports. Then these  values can be updated at a cost $\cO(L)$ per coflow at each iteration. The number of operations required at Lines $11$--$13$ is $\cO(N)$ and at Lines $15$--$16$ is $\cO(N)$, so that finally across iterations it adds to $\cO(N^2)$. On the other hand, it is easy to verify the complexity of $\mathtt{RemoveLateCoflows}$ is  $\cO(NL)$.

\section{Performance Evaluation 
\label{sec:Evaluation}}

We evaluate via simulations our proposed heuristics (two variants of $\mathtt{DCoflow}$\footnote{
The flow-level simulator and the implementation of all algorithms tested in this paper are available at https://github.com/luuquangtrung/CoflowSimulator.}) along with some existing algorithms such as $\mathtt{CS\text{-}MHA}$\footnote{
We only reimplemented the centralized algorithm ($\mathtt{CS\text{-}MHA}$) presented in \cite{Luo2016}, which has been reported to be better than the decentralized version ($\mathtt{D\textsuperscript{2}\text{-}CAS}$) in terms of CAR.
} 
and the solution provided by the optimization method $\mathtt{CDS\text{-}LP}$ proposed in \cite{Tseng2019}. The relaxed version of $\mathtt{CDS\text{-}LP}$, named $\mathtt{CDS\text{-}LPA}$, is also implemented\footnote{
It is worth noting that the formulation of $\mathtt{CDS\text{-}LP}$ and $\mathtt{CDS\text{-}LPA}$ use the same decision variables $\{z_k\}_{k \in \cC}$ as those introduced in Problem~\eqref{prob:CAR}. In $\mathtt{CDS\text{-}LP}$, $z_k$ are binaries, whereas in $\mathtt{CDS\text{-}LPA}$, $z_k$ are continuous variables taking values in $[0,1]$. For any solution yielded by $\mathtt{CDS\text{-}LPA}$, only coflows $k$ whose ${z_k}$ strictly equals $1$ are considered as accepted ones.
}. By using the solution derived from $\mathtt{CDS\text{-}LP}$ as an upper bound, we would get the sense of how close the algorithms are to the optimum. 
A brief description of the reference algorithms has been given in Sec.~\ref{sec:Introduction}. We also compare the performance of our schedulers against $\mathtt{Sincronia}$ \cite{Agarwal2018} and $\mathtt{Varys}$ \cite{Chowdhury2014} that aim to minimize the average CCT.  

Once we obtained the $\sigma$-order, the actual coflow resource allocation for our solution is implemented by the greedy rate allocation algorithm $\mathtt{GreedyFlowScheduling}$ \cite{Agarwal2018}.  At any given point of time, $\mathtt{GreedyFlowScheduling}$ reserves the full port bandwidth to one flow at the time. It does so by complying to the scheduling order in $\sigma$ of the coflow to which the flow belongs \cite{Agarwal2018}. We note that, in the case of $\mathtt{CDS\text{-}LP}$, $\mathtt{CDS\text{-}LPA}$, and $\mathtt{Varys}$, instead, the rate allocation is part of the algorithmic solution. 

The network fabric is represented by $M$ machines or end-hosts connected to a non-blocking Big-Switch fabric, of which each access port has a \textit{normalized} capacity of $1$. The algorithms will be evaluated on both small-scale and large-scale networks, where a network is denoted by $[M, N]$ to indicate different fabric size and number of coflows ($N$) used in the simulations. Small-scale networks have a fabric of size $M = 10$, whereas large-scale networks have a fabric with either $50$ or $100$ machines. The coflows in these networks are generated from both synthetic and real traffic traces.  

The MILP solver $\mathtt{gurobi}$ is used to solved $\mathtt{CDS\text{-}LP}$ and $\mathtt{CDS\text{-}LPA}$. Due to the high complexity, $\mathtt{CDS\text{-}LP}$ and $\mathtt{CDS\text{-}LPA}$ are only evaluated on small-scale networks. In what follows, the detailed setup, comparison metrics, and simulation results are presented.  

\subsection{Simulation Setup
\label{subsec:Simulation-Setup}}

\noindent{\em Synthetic Traffic.}
The synthetic traffic comprises two types of coflows. Type-$1$ coflows have only one flow, whereas the number of flows of Type-$2$ coflows follows a uniform distribution in $[2M/3, M]$. Each generated coflow is randomly assigned to either Class $1$ or Class $2$ with probability of respectively $0.6$ and $0.4$. 
Moreover, each coflow $k$ is assigned a random deadline within $[\text{CCT}_k^0, 2\text{CCT}_k^0]$), where $\text{CCT}_k^0$ is the completion time of coflow $k$ \emph{in isolation}.
Flows of Class-$1$ coflows are assigned a random volume of mean of $1$ and standard deviation of $0.2$. The volume ratio for the flows of Class-$1$ and Class-$2$ coflows is $0.8$.

\vspace{0.2cm}
\noindent{\em Real Traffic.}
Real traffic datasets are obtained by the Facebook traces dataset \cite{Chowdhury2014}, based on a MapReduce shuffle trace collected from one of Facebook's $3000$-machine cluster with $150$ racks. The data traces consist of $526$ coflows. It has a skewed coflow width distribution, ranging from coflows with a single flow to very large ones (the largest coflow has $21170$ flows). For detailed statistics of the Facebook traces we refer the reader to \cite{ChowdhuryThesis2015}. For each configuration $[M,N]$, $N$ coflows are randomly sampled from the Facebook dataset. Coflows are only selected from the ones that have at most $M$ flows. The volume of each flow is given by the dataset.

\vspace{0.2cm}
\noindent{\em Metric.}
We evaluate the algorithms based on the average CAR. We also present  the gains in percentiles of each algorithm with respect to the solution yielded by $\mathtt{CDS\text{-}LP}$ in terms of CAR. These gains are calculated using the formula: $\text{average gain in CAR} = \frac{\text{compared CAR}}{\text{CAR under } \mathtt{CDS\text{-}LP}} - 1.$



\subsection{Results with Offline Setting  \label{sec:Eva:Offline}}

In the offline setting, we consider that all coflows arrive at the same time, i.e., their release time is zero. For each simulation with a specific scale of the network and either synthetic or real traffic traces, we randomly generate $100$ different instances and compute the average performance of algorithms over $100$ runs.

\subsubsection{Average CAR Under Synthetic Traffic}\label{sec:Eva:Offline:Rd}

\begin{figure}[h]
\centering
\subfloat[{Synthetic traffic
traces on a small-scale network.\label{fig:Off:Rate:Random:Small} }]{
\includegraphics[width=.7\columnwidth]{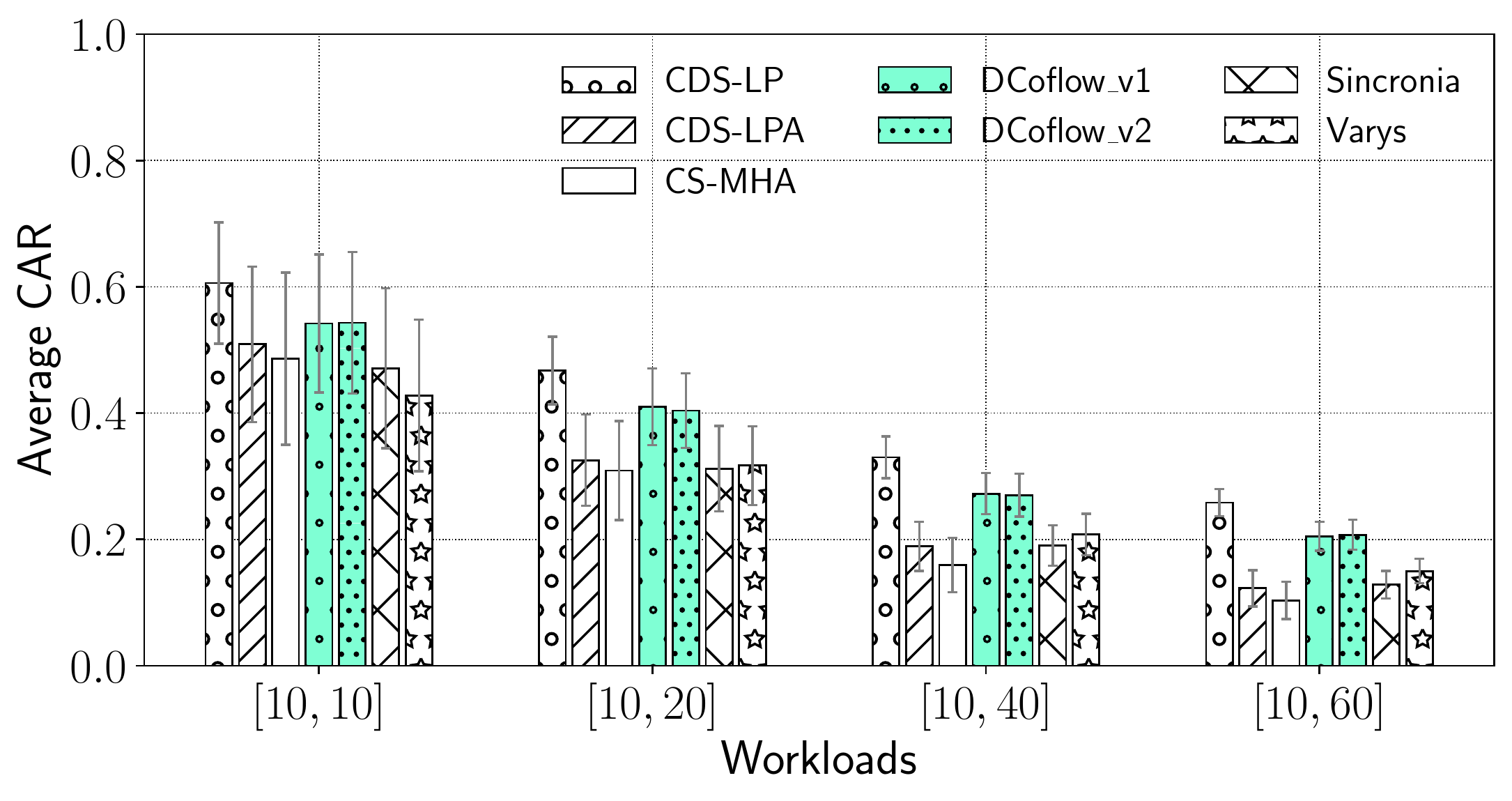}}

\vspace{-0.2cm}

\subfloat[{Synthetic traffic
traces on a large-scale network.\label{fig:Off:Rate:Random:Big} }]{
\includegraphics[width=.7\columnwidth]{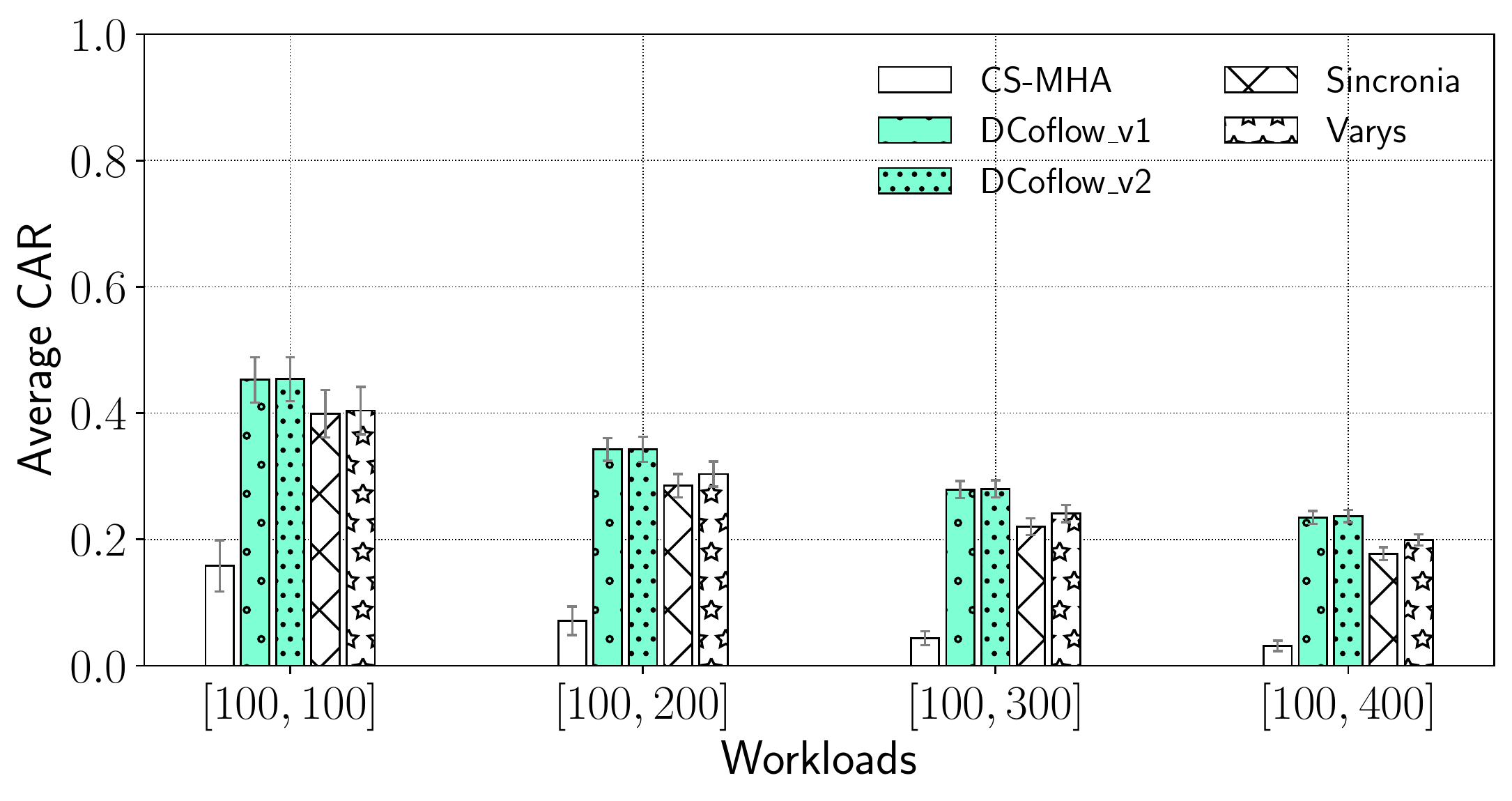}}
\caption{Average CAR with synthetic traffic traces using (a) small-scale and (b) large-scale networks. Each point in the x-axis represents the network $[M,N]$. \label{fig:Off:Rate:Random}}
\end{figure}

Figs.~\ref{fig:Off:Rate:Random:Small}--\ref{fig:Off:Rate:Random:Big} show the average CAR with respectively small-scale networks and large-scale networks. The percentile gains of each algorithm with respect to $\mathtt{CDS\text{-}LP}$  are shown in Fig.~\ref{fig:Off:Percentile:Random}, in terms of average CAR for the configuration $[10,60]$.

It is observed that our proposed heuristics are closest in terms of CAR to the optimum ($\mathtt{CDS\text{-}LP}$) than all other algorithms, with both small- and large-scale networks. Among two variants of $\mathtt{DCoflow}$, $\mathtt{DCoflow\_v1}$ yields the best performance and even outperforms $\mathtt{CDS\text{-}LPA}$, the approximation version of $\mathtt{CDS\text{-}LP}$. 
For instance, on the network $[10,10]$, $\mathtt{DCoflow\_v1}$ improves the CAR on average by 
$6.5\%$,
$11.5\%$,
$15.1\%$, and
$26.6\%$,
compared respectively to 
$\mathtt{CDS\text{-}LPA}$, 
$\mathtt{CS\text{-}MHA}$,
$\mathtt{Sincronia}$, and
$\mathtt{Varys}$. 
Interestingly, the improvement becomes higher when the load is increased. For example, the corresponding improvement on average CAR on a $[10,60]$ network are 
$67.2\%$,
$98.3\%$,
$59.9\%$, and
$36.8\%$ 
(see Fig.~\ref{fig:Off:Rate:Random:Small}). The improvement is even higher when performed on a large-scale netork. For example, compared to 
$\mathtt{CS\text{-}MHA}$,
$\mathtt{Sincronia}$, and
$\mathtt{Varys}$, on the network $[100,400]$, the improvement in terms of average CAR are respectively
$648.1\%$,
$32.3\%$, and
$17.9\%$.
(see Fig.~\ref{fig:Off:Rate:Random:Big}). It is worth noting how the performance of $\mathtt{CS\text{-}MHA}$ falls drastically when dealing with large-scale networks.
This is expected since  $\mathtt{CS\text{-}MHA}$ computes prioritizes coflows that use a large number of ports over those that use a few. In instances with a large number of coflows of the latter type, the CAR of $\mathtt{CS\text{-}MHA}$ goes to $0$ (see detailed explanation with the motivating example in Sec.~\ref{subsec:Motivating-Example})


The result in Fig.~\ref{fig:Off:Percentile:Random} shows that the two variants of $\mathtt{DCoflow}$ achieve a smaller gap to the optimal in almost all values of percentile compared to other algorithms. For instance, compared to $\mathtt{Sincronia}$, $\mathtt{DCoflow\_v1}$ improves the CAR in $50\%$ of $100$ instances by $50\%$ and it achieves around $43\%$ at $99$\textsuperscript{th} percentile. 

\subsubsection{Average CAR Under Real Traffic Traces}\label{sec:Eva:Offline:Fb}

\begin{figure}[t]
\centering
\subfloat[{Facebook traffic traces on a small-scale network.\label{fig:Off:Rate:Facebook:Small} }]{
\includegraphics[width=.7\columnwidth]{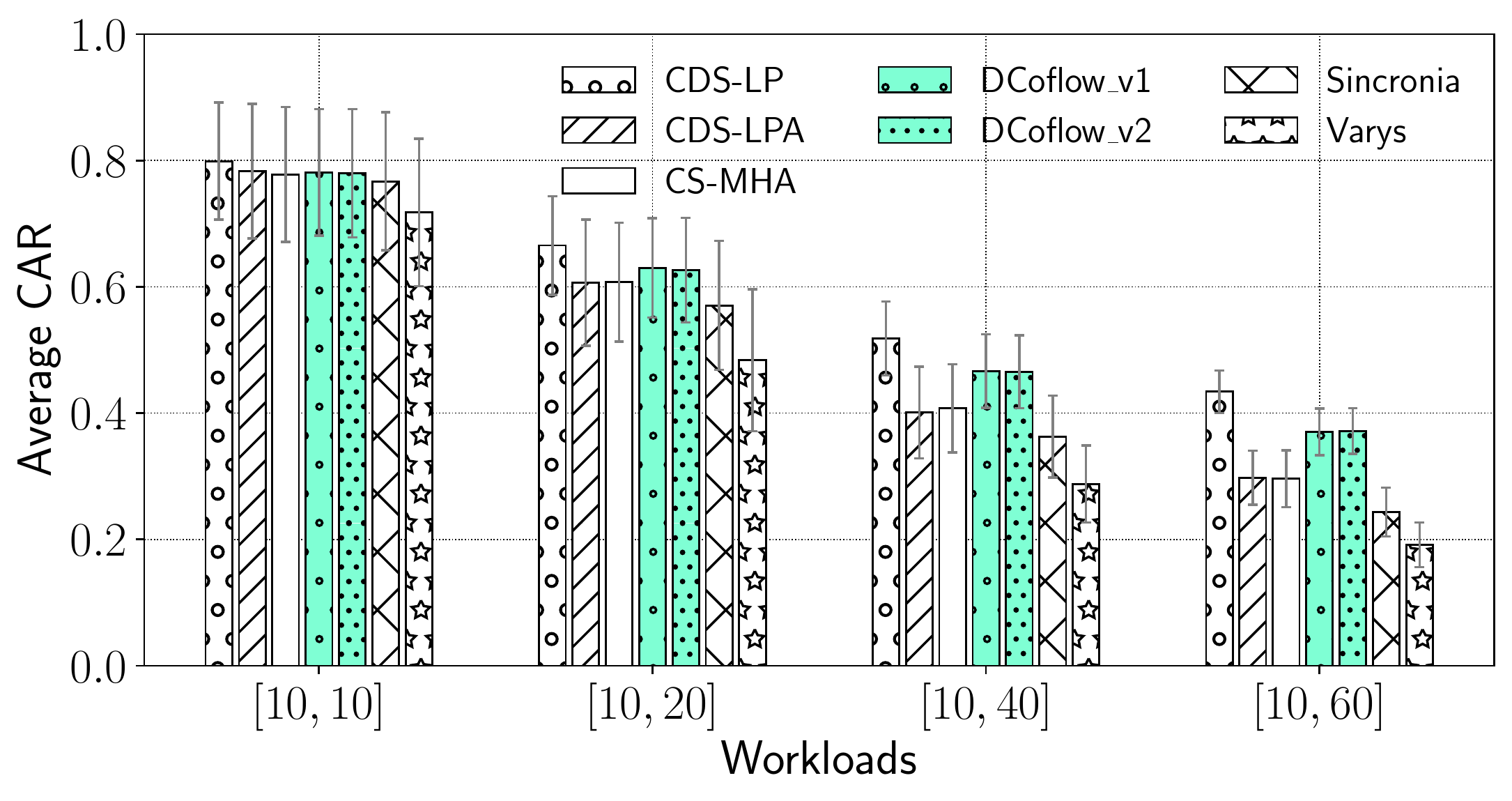}}

\vspace{-0.2cm}

\subfloat[{Facebook traffic traces on a large-scale network.\label{fig:Off:Rate:Facebook:Big} }]{
\includegraphics[width=.7\columnwidth]{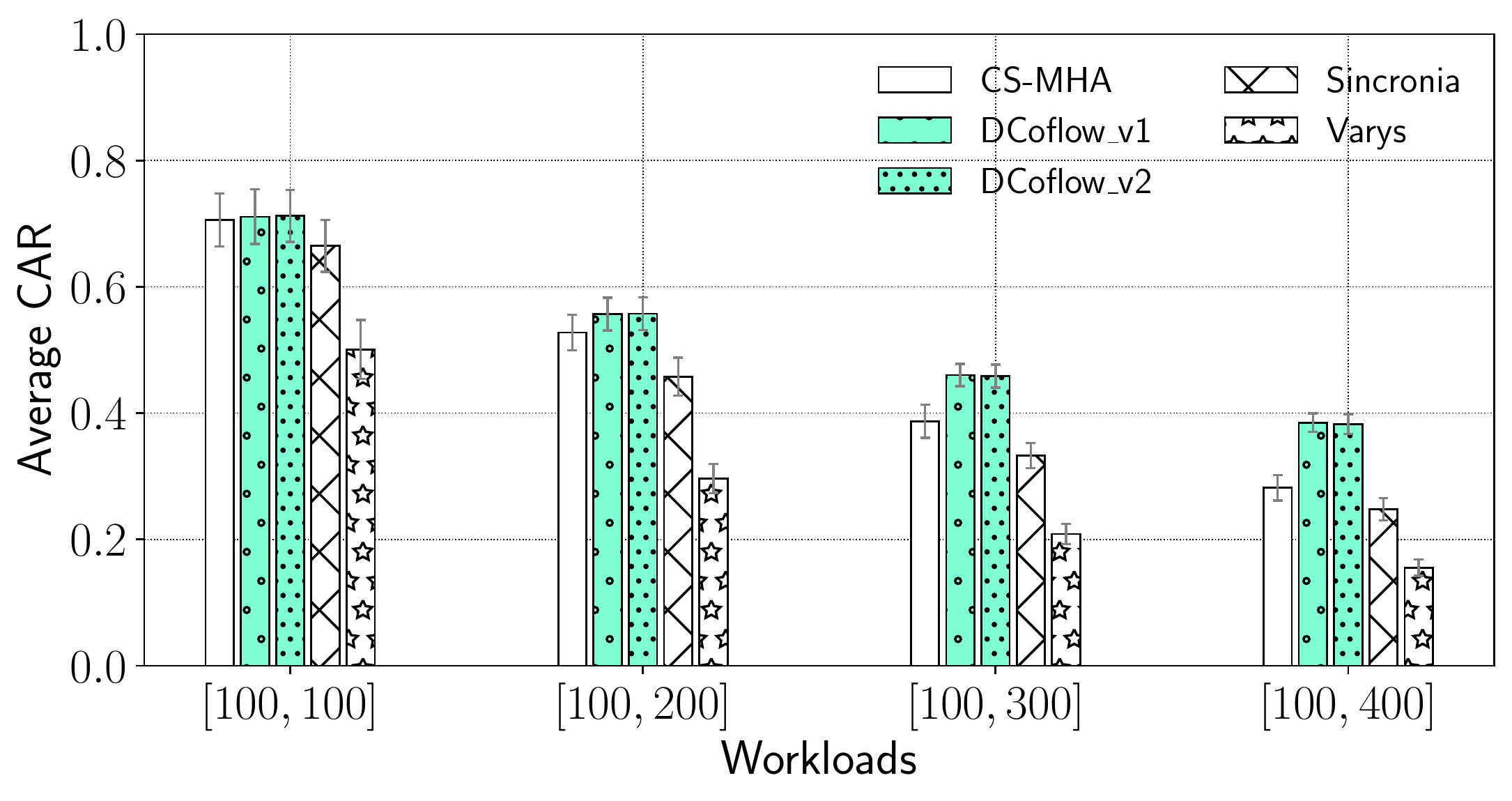}}
\caption{Average CAR with Facebook traces using (a) small-scale network and (b) large-scale network. Each point in the x-axis represents network $[M,N]$. \label{fig:Off:Rate:Facebook}}
\end{figure}

\begin{figure}[t]
\centering
\subfloat[{Synthetic traces. \label{fig:Off:Percentile:Random}}]{
\includegraphics[width=0.4\columnwidth]{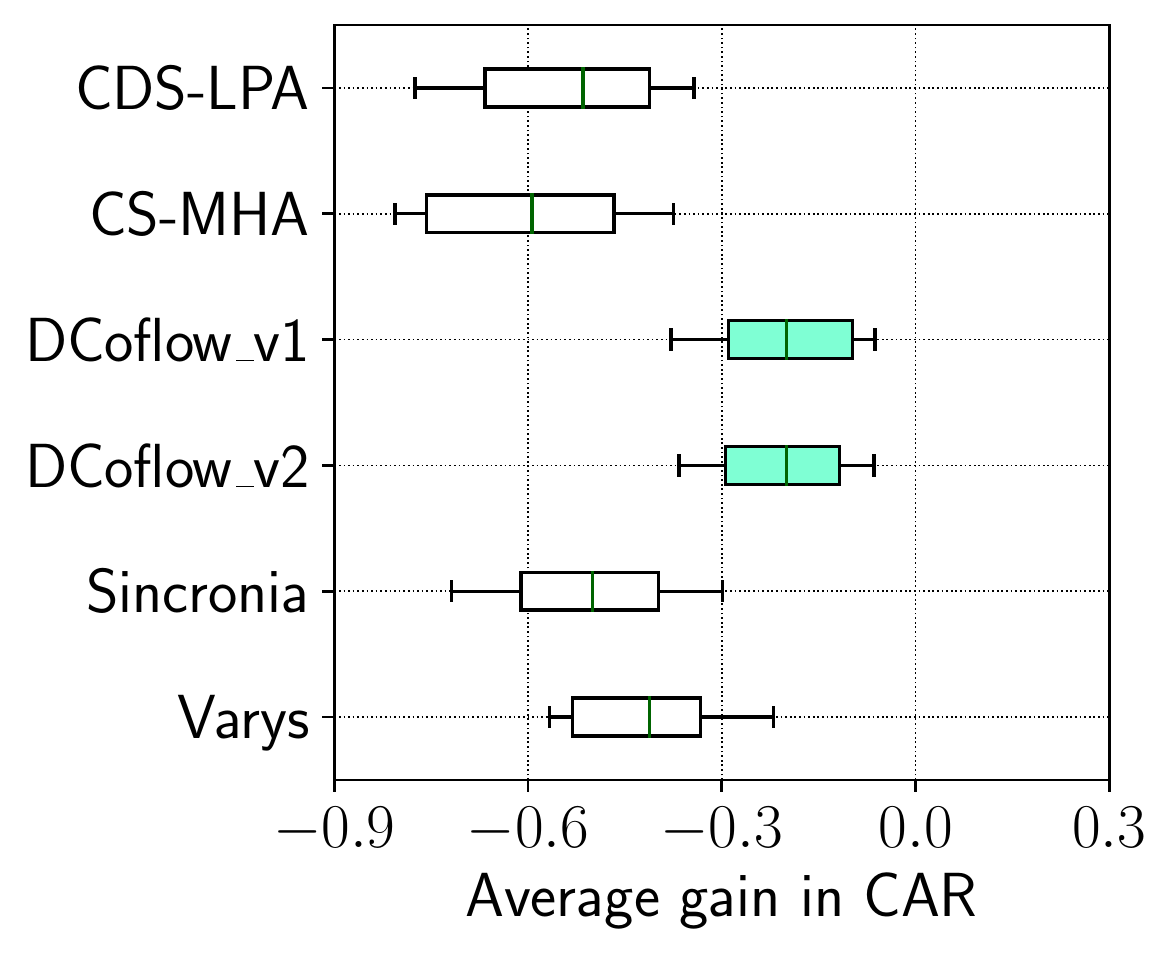}}
\subfloat[{Facebook traces.\label{fig:Off:Percentile:Facebook} }]{
\includegraphics[width=0.4\columnwidth]{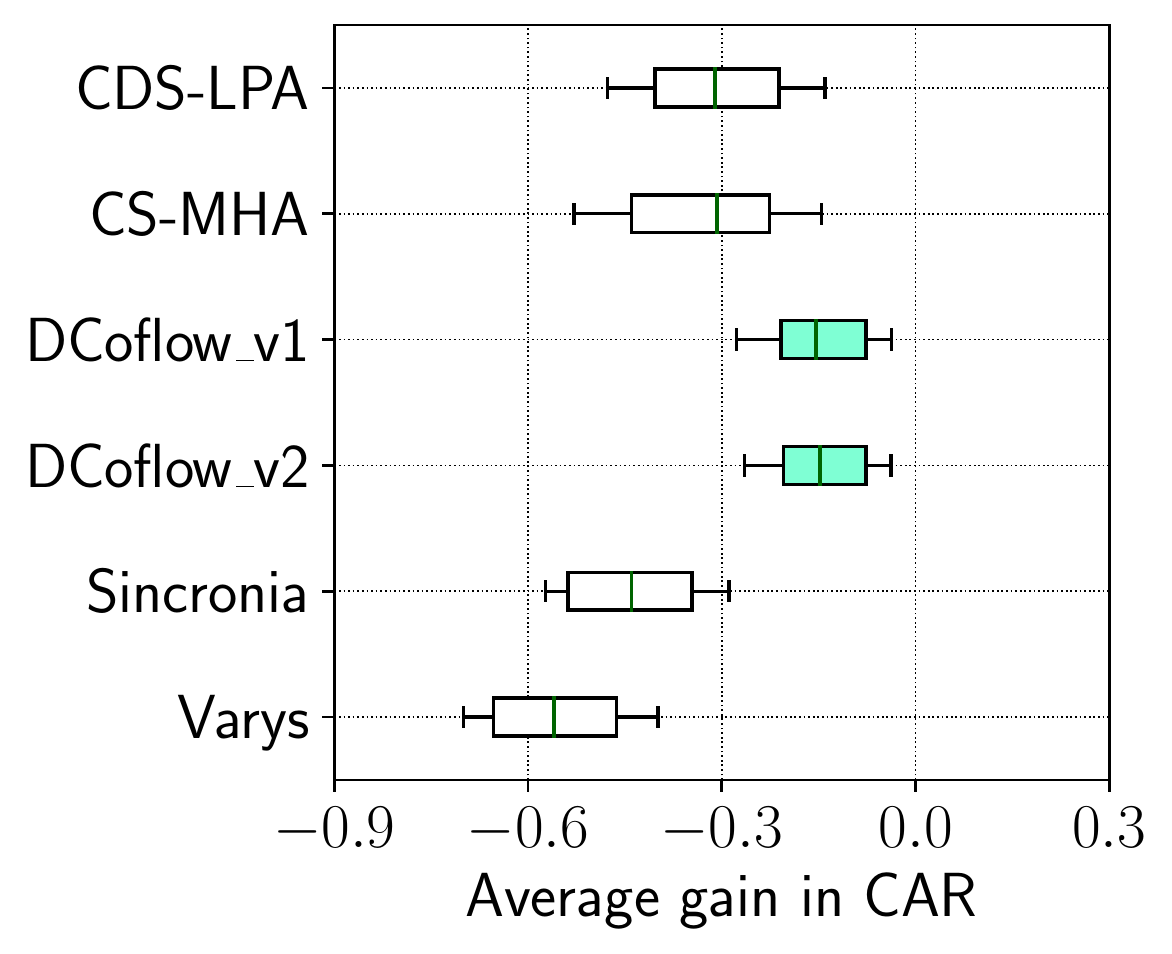}}
\caption{The $1$\textsuperscript{st}-$10$\textsuperscript{th} -$50$\textsuperscript{th}-$90$\textsuperscript{th}-$99$\textsuperscript{th} percentiles of the average gain in CAR with small-scale network $[10,60]$ using (a) synthetic and (b) Facebook traces. \label{fig:Off:Percentile}}
\vspace{-0.5cm}
\end{figure}

This section presents the results obtained with the Facebook traffic traces, using the same configurations as those used in Sec.~\ref{sec:Eva:Offline:Rd}. Figs.~\ref{fig:Off:Rate:Facebook:Small}--\ref{fig:Off:Rate:Facebook:Big} show the average CAR with respectively small- and large-scale networks. The gains in percentiles of each algorithm with respect to $\mathtt{CDS\text{-}LP}$, in terms of average CAR when using a $[10,60]$ fabric are shown in Fig.~\ref{fig:Off:Percentile:Facebook}. Similar to what observed in the results with the synthetic traces (see Sec.~\ref{sec:Eva:Offline:Rd}), $\mathtt{DCoflow\_v1}$ and $\mathtt{DCoflow\_v2}$ yield a significant improvement in terms of average CAR compared to other heuristics. For instance, with a $[10,60]$ configuration, the two variants of $\mathtt{DCoflow}$ improve the average CAR on average by  
$24.4\%$,
$25\%$,
$52.2\%$,
$93.1\%$,
compared respectively to 
$\mathtt{CDS\text{-}LPA}$, 
$\mathtt{CS\text{-}MHA}$,
$\mathtt{Sincronia}$, and
$\mathtt{Varys}$.
(see Fig.~\ref{fig:Off:Rate:Facebook:Small}). The improvement is even higher when performed on a large-scale network. For example, compared to 
$\mathtt{CS\text{-}MHA}$,
$\mathtt{Sincronia}$, and
$\mathtt{Varys}$, on $[100,400]$ network, the improvement in terms of average CAR are respectively
$36.6\%$,
$55.3\%$, and
$147.5\%$.

Moreover, the results in Fig.~\ref{fig:Off:Percentile:Facebook} show that the two variants of $\mathtt{DCoflow}$ achieve a smaller gap to the optimal in almost all values of percentile compared to the other algorithms. 
For instance, compared to $\mathtt{Sincronia}$, $\mathtt{DCoflow\_v1}$ improves the CAR in $57\%$ of $100$ instances by $50\%$ and it achieves around $35\%$ at $99$\textsuperscript{th} percentile. 

\subsubsection{Prediction Error of $\mathtt{DCoflow}$}

As mentioned at the end of Sec.~\ref{sec:heuristic}, we also evaluate the prediction error $(|\sigma| - |\hat{\sigma}|)/|\sigma|$ of our heuristics, where $\hat{\sigma} \subseteq \sigma$ is the set of coflows in $\sigma$ that satisfy the deadline constraint after performing the actual resource allocation using $\mathtt{GreedyFlowScheduling}$. Both variants of $\mathtt{DCoflow}$ provide a prediction of CAR with an average error below $3.6\%$ for both traffic traces.

\subsection{Online Setting 
\label{sec:Eva:Online}}

We now present a series of numerical results on the performance of the online version of $\mathtt{DCoflow}$. The metric used for the performance evaluation is the average CAR obtained over $40$ instances. In each instance, coflows arrive sequentially according to a Poisson process of rate $\lambda$, i.e., the inter-arrival time of coflows is exponentially distributed with rate $\lambda$. Unless stated otherwise, 
coflow priorities 
are computed  upon arrival of a new coflow ($f=\infty$). 

As both versions of $\mathtt{DCoflow}$ provide similar results, we only present the results obtained with $\mathtt{DCoflow\_v1}$. The average CAR obtained with $\mathtt{DCoflow\_v1}$ is compared against those obtained with the online version of Varys with deadline \cite{Ma2016}, $\mathtt{CS\text{-}MHA}$ and $\mathtt{Sincronia}$. We investigate the effect of two main parameters: (\textit{i}) the coflow arrival rate $\lambda$ and (\textit{ii}) the frequency $f$ at which coflow priorities are updated.


\subsubsection{Impact of Arrival Rate}
We first study the impact of the arrival rate $\lambda$ on the CAR obtained with the various algorithms. The CAR is averaged over $40$ instances, each one with $4000$ coflow arrivals. %
The deadline of a coflow $k$ is drawn from a uniform distribution in $[\text{CCT}_k^0, 4\text{CCT}_k^0]$. We also consider two scenarios: a small fabric with $M = 10$ machines, and a large fabric with $M = 50$ machines. For each scenario, results are presented for the following values of $\lambda$: $\lambda = 8$, $\lambda = 12$, $\lambda = 16$, and $\lambda = 20$. 

\begin{figure}[t]
\centering
\subfloat[{Small fabric.\label{fig:On:Rate:Random:Rachid:M10} }]{
\includegraphics[width=.49\columnwidth]{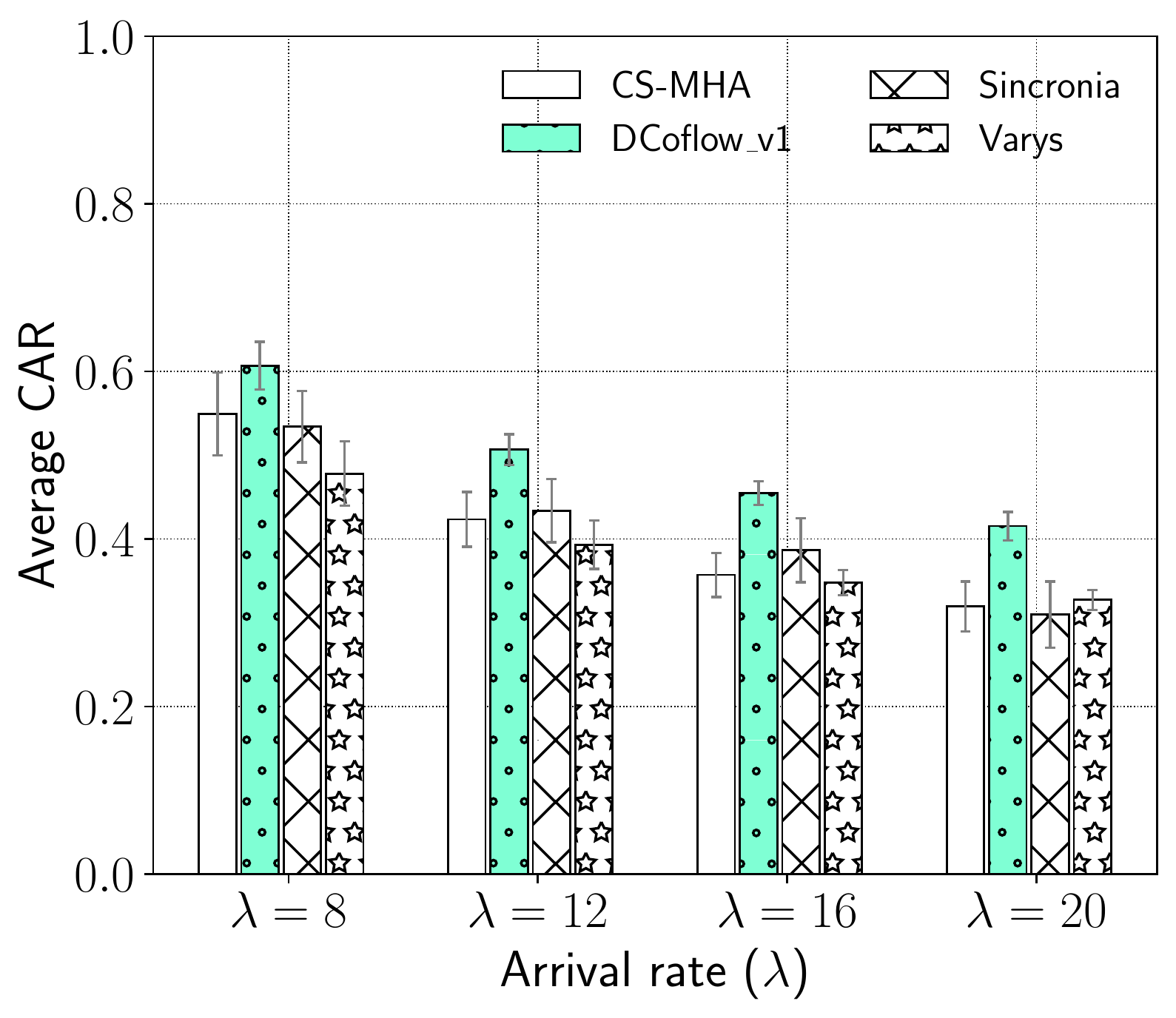}}
\subfloat[{Large fabric.\label{fig:On:Rate:Random:Rachid:M50} }]{
\includegraphics[width=.49\columnwidth]{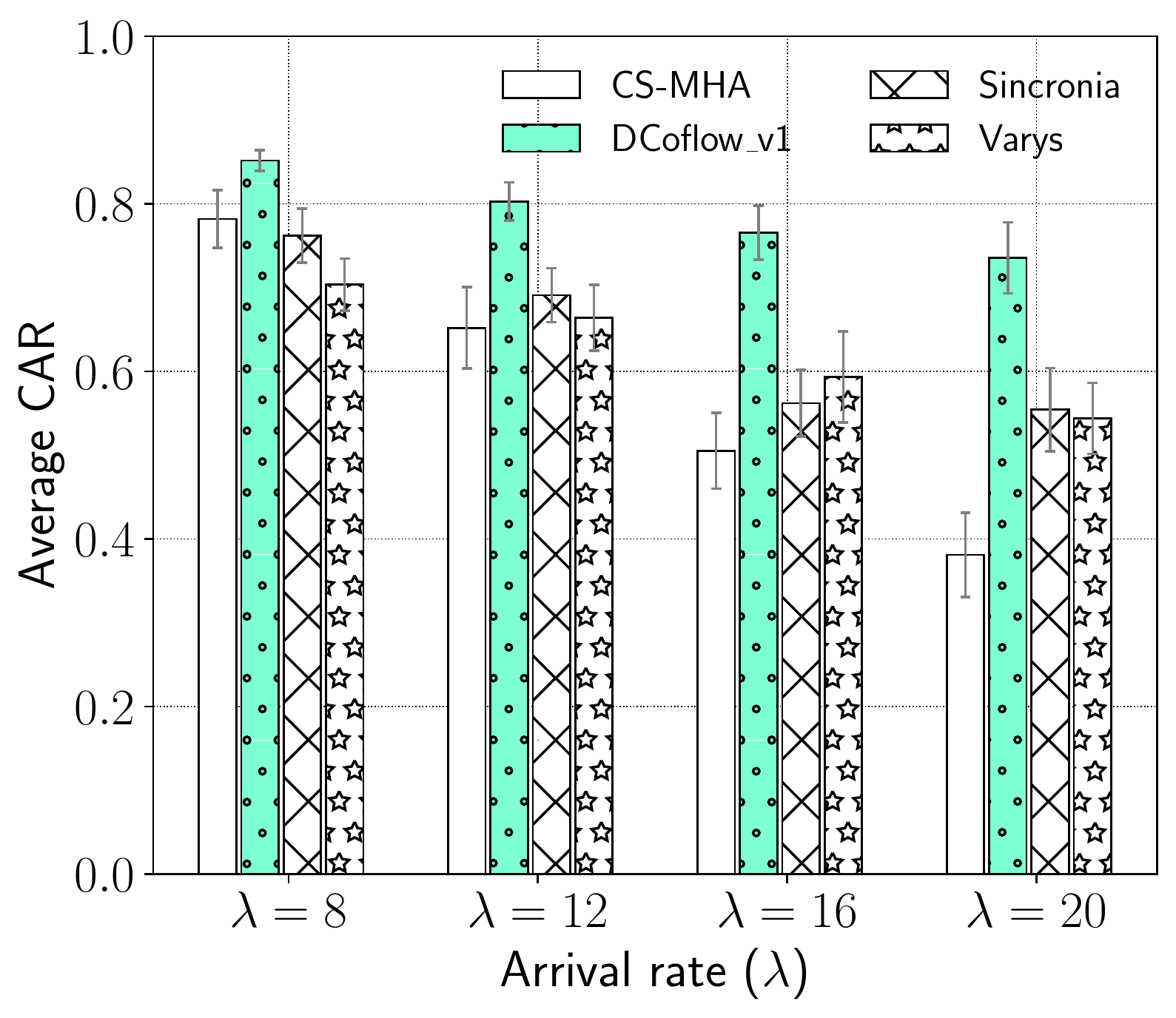}}
\caption{Average CAR using synthetic traffic with varying $\lambda$ and (a) $M = 10$ and (b) $M = 50$. \label{fig:On:Rate:Random:Rachid}}
\end{figure}

Our results are shown in Figs.~\ref{fig:On:Rate:Random:Rachid:M10} and \ref{fig:On:Rate:Random:Rachid:M50}, for the small fabric scenario and the large one, respectively. We observe that $\mathtt{DCoflow\_v1}$ achieves a higher average CAR for all values of $\lambda$, and that the gain with respect to the other scheduling algorithms increases with the value of $\lambda$. If all algorithms achieve more or less the same CAR for a lightly loaded fabric, $\mathtt{DCoflow\_v1}$ clearly outperforms the other algorithms when the fabric is highly congested.

\begin{figure}[htb]
\centering
\subfloat[{Small fabric.\label{fig:On:Rate:Facebook:Rachid:M10C4000F60} }]{
\includegraphics[width=.49\columnwidth]{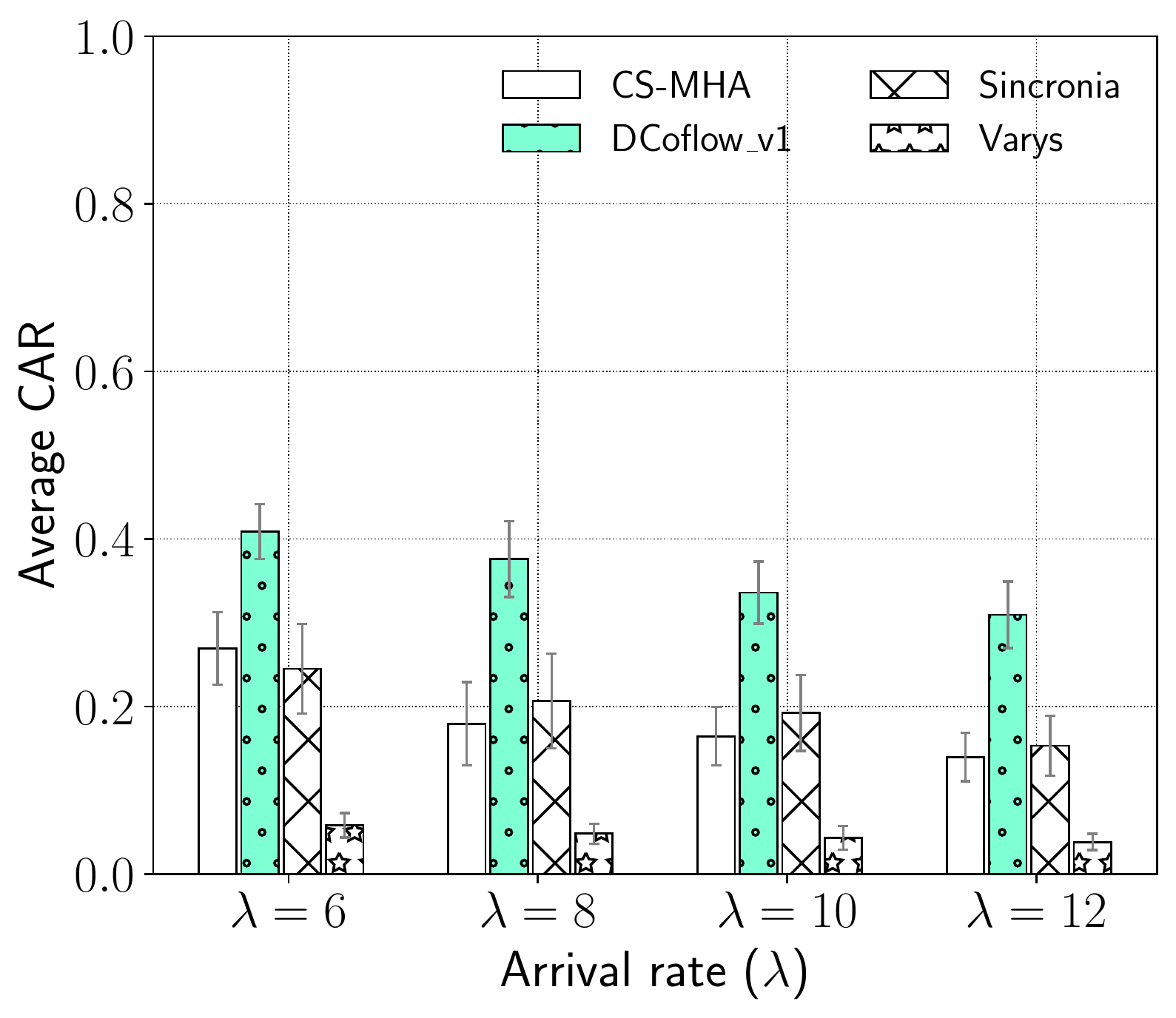}}
\subfloat[{Large fabric.\label{fig:On:Rate:Facebook:Rachid:M100C4000F200} }]{
\includegraphics[width=.49\columnwidth]{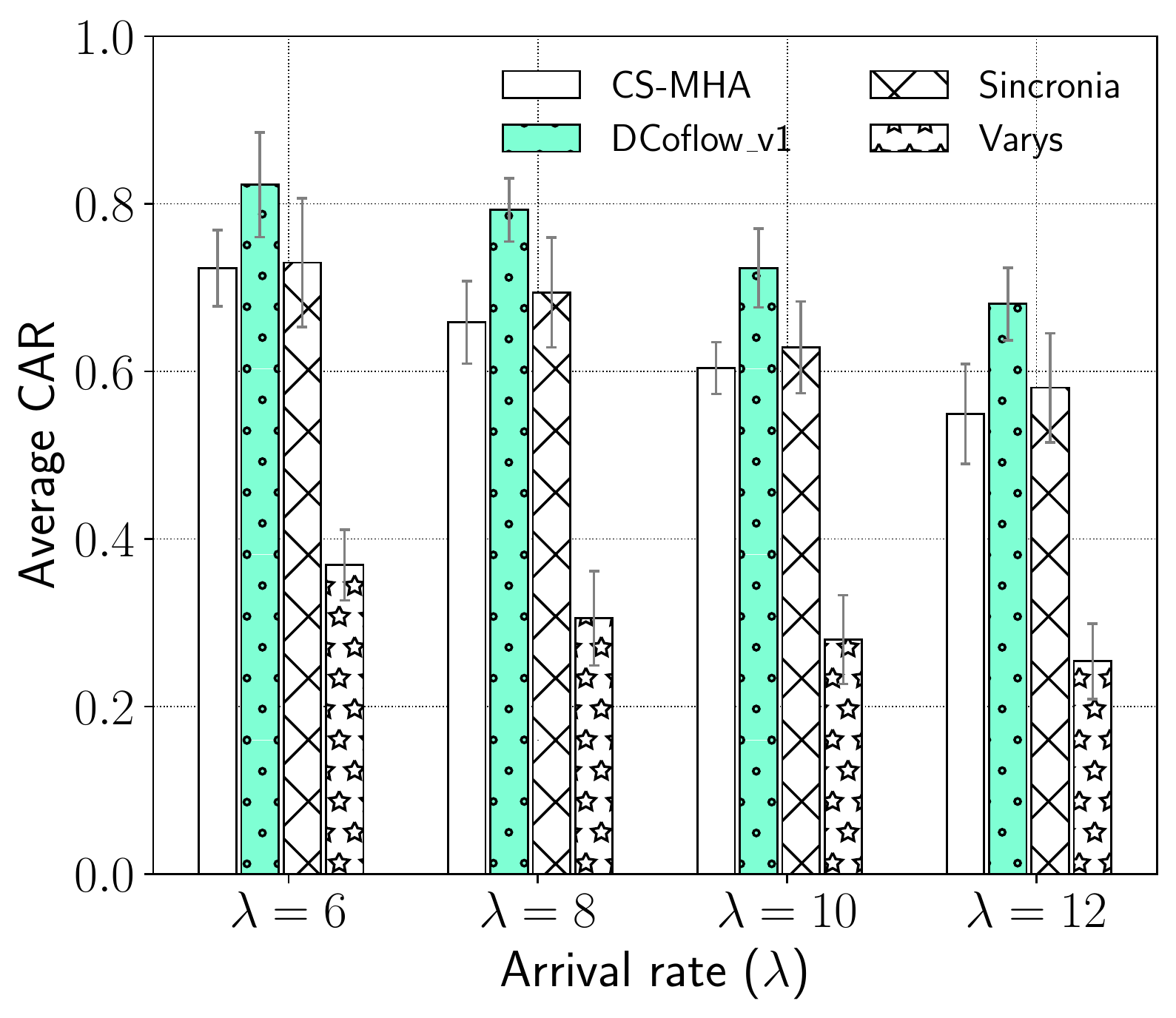}}
\caption{Average CAR using Facebook traffic with varying $\lambda$ and (a) $M = 10$ and (b) $M = 100$. \label{fig:On:Rate:Facebook:Rachid}}
\end{figure}

Figs.~\ref{fig:On:Rate:Facebook:Rachid:M10C4000F60} and \ref{fig:On:Rate:Facebook:Rachid:M100C4000F200} show respectively the average CAR obtained with $M=10$ and $M=100$, both with $4000$ coflows, using the Facebook dataset. Similar to what obtained with the synthetic traffic traces, $\mathtt{DCoflow\_v1}$ outperforms all other methods with significant gains. When the fabric is highly congested (i.e., with $M=10$), again $\mathtt{DCoflow\_v1}$ yields a higher gain compared to other algorithms. For instance, $\mathtt{DCoflow\_v1}$ achieves $9.3\%$ higher CAR than $\mathtt{Sincronia}$ when $M=100$ (see Fig.~\ref{fig:On:Rate:Facebook:Rachid:M100C4000F200}), while with $M=10$, the gap becomes $16.4\%$ (see Fig.~\ref{fig:On:Rate:Facebook:Rachid:M10C4000F60}).

\subsubsection{Impact of Update Frequency}


We now evaluate the impact of the update frequency $f$ on the average CAR. We consider the following values of $f$: $f=\frac{\lambda}{2}$, $f=\lambda$, $f=2\lambda$, and $f=\infty$. Recall that $f=\infty$ means that priorities are updated upon each coflow arrival. We assume that $M=10$ and compute the CAR by averaging over $40$ instances. For each instance, we simulate $8,000$ coflow arrivals according to a Poisson process at rate $\lambda$, assuming that the deadline of a coflow $k$ is uniformly distributed in $[\text{CCT}_k^0, 2\text{CCT}_k^0]$. We present the average CAR obtained for different values of $f$ ($f \in \{ \frac{\lambda}{2}, \lambda, 2\lambda,  \infty \}$) and for different values of the arrival rate $\lambda$ (from $2$ to $10$).



\begin{figure}[t]
\centering
\subfloat[{Without batch.\label{fig:On:Rate:Random:Olivier:nobatch} }]{
\includegraphics[width=.49\columnwidth]{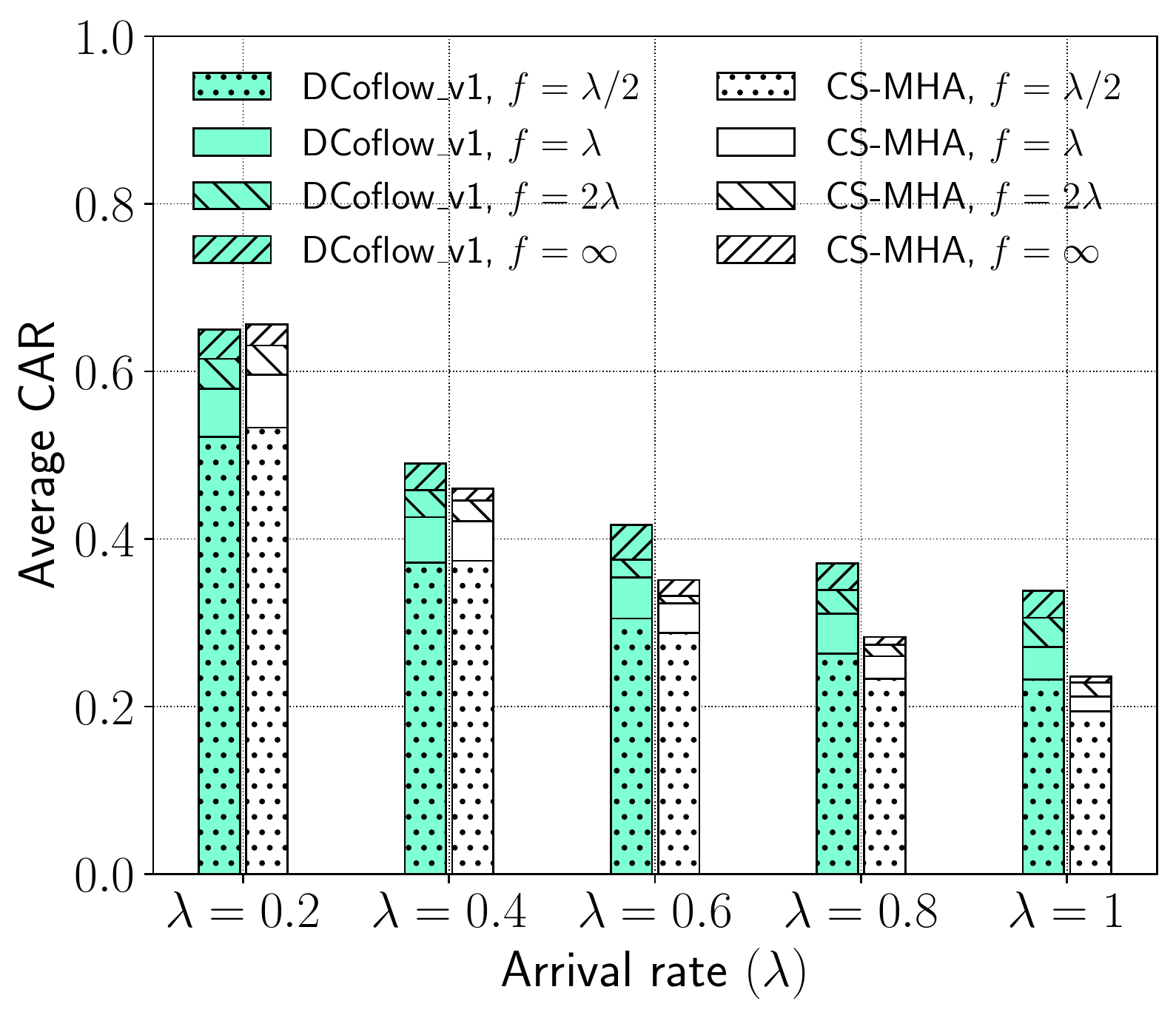}}
\subfloat[{With batch.\label{fig:On:Rate:Random:Olivier:batch} }]{
\includegraphics[width=.49\columnwidth]{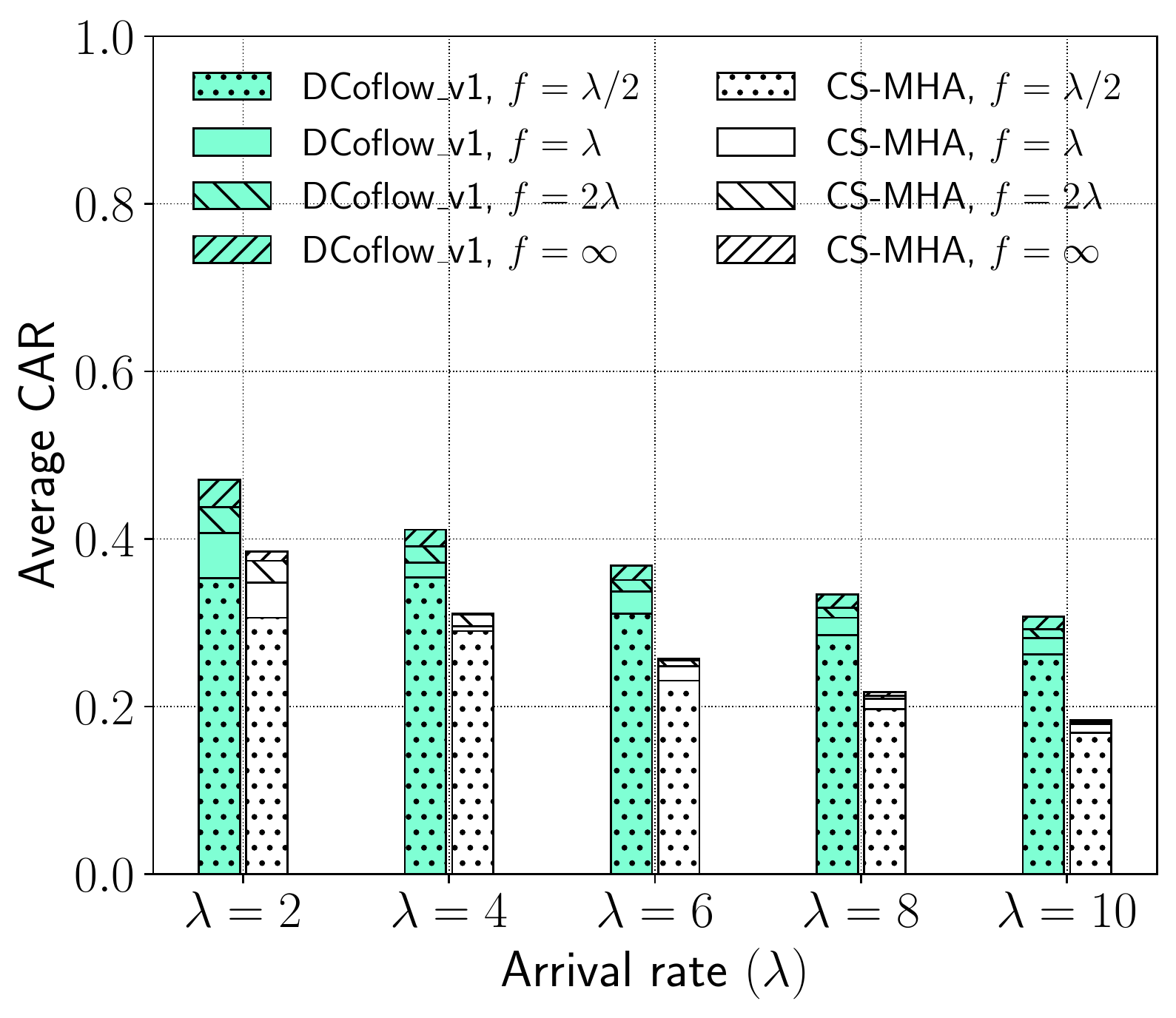}}
\caption{Average CAR of $\mathtt{DCoflow\_v1}$ and $\mathtt{CS\text{-}MHA}$ using synthetic traffic with $[10,8000]$ and varying $\lambda$, when obtaining (a) one single coflow per arrival; and (b) a random batch of coflow per arrival. \label{fig:On:Rate:Random:Olivier}}
\end{figure}

The results in Fig.~\ref{fig:On:Rate:Random:Olivier:nobatch} are obtained from a simulation, in which each arrival corresponds to one single coflow. We note, as before, that for a low arrival rate, both algorithms provide a similar average CAR (for $\lambda=2$, $\mathtt{CS\text{-}MHA}$ achieves a slightly higher CAR than $\mathtt{DCoflow\_v1}$), but that $\mathtt{DCoflow\_v1}$ clearly outperforms $\mathtt{CS\text{-}MHA}$ when the fabric is highly congested. We also note that a higher frequency $f$ significantly improves the CAR for both algorithms. For instance, for $\lambda=2$ (resp. $\lambda=10$), the average CAR is increased by $52\%$ (resp. $46\%$) if we update coflow priorities upon arrival of each new coflow instead of using the periodic scheme with $f=\frac{\lambda}{2}$. These results suggest that there is a need for a trade-off between the computational complexity of updating coflow priorities at a high frequency and the CAR achieved.
In Fig.~\ref{fig:On:Rate:Random:Olivier:batch}, we present a similar result, but assuming that coflows arrive in batches. The batch size follows a uniform distribution $\mathcal{U}([5,15])$. Since the average number of coflows in each batch is $10$, in this setting, we divide the batch arrival rate by $10$ to obtain the same coflow arrival rates as in Fig.~\ref{fig:On:Rate:Random:Olivier:nobatch}. The results obtained for batch arrivals are similar to those obtained previously, but we note that the gains of $\mathtt{DCoflow\_v1}$ with respect to $\mathtt{CS\text{-}MHA}$ are significantly higher in this case. Moreover, we note that the benefit of using a higher update frequency is lower in this case (e.g., for $\lambda=10$, the average CAR is increased only by $17\%$ if we use $f=\infty$ instead of $f=\frac{\lambda}{2}$).

\section{Related Work 
\label{sec:Related-Work} }

As discussed in Sec.~\ref{sec:Introduction}, most works in the literature focus on CCT minimization, and deadline scheduling has received comparatively less attention. $\mathtt{Varys}$ \cite{Chowdhury2014} was one of the first algorithms for deadline-sensitive coflow scheduling. It uses a cascade of coflow admission control and scheduling. The scheduler strives for CCT minimization combining (\textit{i}) a coflow ordering heuristic based on the per coflows bottleneck's completion time; and (\textit{ii}) an allocation algorithm to assign bandwidth to individual flows of each coflow. Rate allocation in $\mathtt{Varys}$ is performed to approximately align the completion time of all coflows to the bottleneck one.


$\mathtt{Chronos}$ \cite{Ma2016} is a heuristic for deadline scheduling which avoids starvation for flows that do not meet their deadlines by granting them the residual bandwidth. A priority order is determined first, and coflows are hence allocated the minimum necessary bandwidth to meet their individual deadlines. 
Once all the flows that meet their deadlines have been allocated bandwidth, the residual bandwidth is shared by the remaining coflows in proportion to their demands. 

In \cite{Luo2016}, a connection between deadline scheduling of coflows and the well-known problem of minimizing late jobs in a concurrent open shop---a known \textit{NP}-hard problem, is made. A heuristic based on the Moore-Hodgson's algorithm \cite{Moore1968} for a single link is proposed. Both centralized as well as decentralized heuristics are introduced (namely $\mathtt{CS\text{-}MHA}$ and $\mathtt{D\textsuperscript{2}\text{-}CAS}$, respectively). 

A formal description of the deadline scheduling problem including bandwidth allocation of flows was given in \cite{Tseng2019}. The CDS maximization problem is formulated as an MILP (called $\mathtt{CDS\text{-}LP}$). Time is divided into intervals whose boundaries are the coflow deadlines arranged in increasing order. The program determines which coflows to accept and the amount of bandwidth to allocate in each interval. $\mathtt{CDS\text{-}LP}$ is shown to be \textit{NP}-hard, and an approximation based on LP relaxation (called $\mathtt{CDS\text{-}LPA}$) of the binary variables is also proposed. $\mathtt{CDS\text{-}LPA}$ only accepts coflows for which the relaxed variable is strictly equal to $1$, i.e., only coflows that are completely accepted by the LP relaxation are retained. 

For completeness, we also cite a few works considering the minimization of CCT \cite{chowdhury2015coflow,agarwal2018sincronia,Chen2016,Shaf2018,Shi2021} as well as the survey article \cite{Wang2018}. Popular among CCT minimization algorithms is $\mathtt{Sincronia}$ proposed in \cite{Agarwal2018}. It considers scheduling on the network bottlenecks and returns a scheduling coflow order achieving a $4$--approximation factor. 

\section{Conclusion and Future Work
\label{sec:Conclusion}}

In this paper, we introduced a new joint coflow admission control and scheduling algorithm for a batch of coflows with deadlines. The proposed schemes leverage results from open-shop scheduling to determine a subset of coflows to schedule and a corresponding  $\sigma$-order which is then employed in order to schedule coflows in priority. 

Numerical results show that on small-scale networks, our algorithms perform similar to or better than other deadline-sensitive algorithms proposed in the literature. On large-scale networks, however, it shows significant improvements with respect to existing algorithms, e.g., $98\%$ higher CAR than $\mathtt{CS\text{-}MHA}$ in an offline setting. Our scheme also has a low prediction error: even though the admission control is performed using a bottleneck approximation for the CCT, almost all accepted coflows actually finish within their deadline when being actually scheduled. 

This behaviour is observed in both offline and online settings, both with synthetic traces and for real traces from the Facebook data set. This shows that the algorithm is robust with respect to the coflow size distribution and performs very well across a wide range of network sizes. 

Several extensions of this research line are possible. In future works, we shall study the performance of our online solution in the case when coflows tend to be released in batches, e.g., when a distributed framework polls worker nodes with given period. Furthermore, a relevant case is that of incomplete information on flow volumes, i.e., because the volume of a flow is not directly available to the scheduler but only, for instance, via some \textit{a priori} distribution. Finally, issues of starvation and fairness issue among coflows represent interesting issues we have not addressed yet.


{\scriptsize
\bibliographystyle{IEEEtran} 
\bibliography{ref_coflow}
}

\end{document}